# HAPS Communication Networks: A Tutorial-cum-Survey on Integration with Optical Atmospheric Sensing

Ali Elkhazraji, Mohamed-Slim Alouini, *Fellow, IEEE*, and Aamir Farooq

*Abstract*—High-Altitude Platform Stations (HAPS) are emerging as a critical component of future non-terrestrial networks (NTNs), capable of supporting gigabit-class free-space optical (FSO) backhaul links while simultaneously hosting advanced laser-based sensing payloads. This tutorial-cum-survey article reviews and synthesizes recent developments in HAPS-based optical communication and provides a tutorial on integration with atmospheric remote sensing using shared optical links. Multiple sensing modalities suitable for HAPS deployment are assessed, among which Differential Absorption Lidar (DIAL) is identified as the most promising owing to its range-resolved sensitivity, spectral selectivity, and strong compatibility with HAPS payload constraints. A comprehensive technical roadmap is presented for implementing telecom-band DIAL on HAPS platforms in coexistence with high-throughput FSO communication systems. The article examines the architectural, spectral, and atmospheric advantages of HAPS over terrestrial and satellite systems, highlighting their superior spatial–temporal coverage, station-keeping capability, and suitability for compact laser payloads such as DIAL, in-situ sensors, and multispectral imagers. Special emphasis is placed on the feasibility of co-locating sensing and communication functions within a shared optical and power envelope, particularly in the telecom C-band (1.53–1.57 μm), enabling simultaneous retrieval of trace gases ($CO_2$, $CH_4$, $N_2O$, $H_2S$, $O_3$) while sustaining multi-Gbps data downlinks. The paper also categorizes suitable HAPS architectures (balloons, fixed-wing UAVs, and airships), addresses platform-level constraints, and identifies operational use cases where integrated sensing and communication (ISAC)-enabled HAPS can outperform satellites and UAVs — including greenhouse gas monitoring, disaster response, urban air-quality mapping, and 6G NTN extensions. Furthermore, a scientometric analysis of 2005–2025 literature reveals that HAPS-related publications have tripled since 2014, underscoring rapid growth in multidisciplinary research interest. The findings confirm that telecom-grade optical hardware, favorable atmospheric transmission windows, and an accelerating research ecosystem are converging to position HAPS as a persistent stratospheric layer for high-capacity communication and environmental observation within the 6G ISAC landscape.

## I. INTRODUCTION

SMART cities are rapidly evolving urban spaces that utilize information and communication technologies to enhance the quality of life for citizens, optimize resource consumption, and improve public service delivery [1]. Connectivity, particularly through wireless technologies, forms the backbone of these smart infrastructures, enabling efficient data exchange between various devices and systems, such as smart grids, transportation networks, and internet of things (IoT)-enabled sensors [2]. High-altitude platform stations (HAPS) are a promising addition to this ecosystem by providing scalable, flexible, and low-latency communication networks that can bridge the connectivity gaps in urban and rural areas [3–6]. These stratospheric vehicles (such as balloons, airships, or unmanned aerial systems) operate between 17 and 22 kilometers above the Earth's surface [7]. HAPS are designed to remain in the air for extended periods, providing communication, surveillance, and remote sensing capabilities over large areas. They can support both short-range and long-range wireless communication technologies, such as 5G and beyond, ensuring that smart cities have a resilient and adaptive communication framework to manage the data-intensive nature of their operations [8,9].

HAPS can also assist in disaster management and public safety operations within smart cities by maintaining uninterrupted communication when terrestrial networks are compromised. The integration of HAPS into smart city infrastructures is seen as a viable solution for overcoming challenges related to urban congestion, increased demand for bandwidth, and connectivity in underdeveloped or hard-to-reach areas [10]. As cities become increasingly digitized, the role of HAPS in providing sustainable and inclusive communication platforms will be pivotal in driving the future of smart city innovations [11]. In rural and underserved regions, HAPS can provide essential communication services where terrestrial infrastructure is absent. Additionally, they offer a flexible alternative to satellites in terms of upgrading and maintaining communication technologies over time [12]. Non-terrestrial networks (NTN), encompassing both satellite and HAPS-based communication systems, are gaining increasing

Ali Elkhazraji is with the Department of Mechanical Engineering, King Fahd University of Petroleum & Minerals, Dhahran 31261, Saudi Arabia (e-mail: ali.elkhazraji@kfupm.edu.sa).

Mohamed-Slim Alouini is with the Computer, Electrical and Mathematical Sciences and Engineering (CEMSE) Division, King Abdullah University of Science and Technology (KAUST), Thuwal 23955-6900, Saudi Arabia (e-mail: slim.alouini@kaust.edu.sa).

Aamir Farooq is with the Physical Science and Engineering Division, King Abdullah University of Science and Technology (KAUST), Thuwal 23955, Saudi Arabia (e-mail: aamir.farooq@kaust.edu.sa).

attention due to their potential to extend network coverage to areas where terrestrial infrastructure is not economically viable [8]. HAPS, specifically, are emerging as a cost-effective and scalable solution for delivering broadband connectivity to remote and underserved regions [13]. Their ability to cover wide geographical areas makes them ideal for providing last-mile connectivity, particularly in developing nations or regions affected by natural disasters [14].

Beyond communications, HAPS are increasingly being explored as persistent aerial nodes for Earth observation, environmental monitoring, and remote sensing. The extended dwell time and high vantage point of stratospheric platforms make them ideal for imaging and atmospheric sensing applications using both optical and non-optical modalities. For example, recent developments in HAPS-based radar instruments, such as the DLR's HAPSAR synthetic aperture radar (SAR), have demonstrated the feasibility of high-resolution radar imaging from the stratosphere [15–18]. Similarly, camera-based payloads have enabled large-area, real-time imaging and surveillance applications from HAPS altitudes [19,20]. These imaging systems complement emerging optical sensing and laser-based technologies, such as optical communication and atmospheric spectroscopy, by providing multi-modal data acquisition capabilities from a shared high-altitude platform.

Remote gas sensing has garnered significant attention [21], motivated by its relevance to greenhouse gas emissions [22], rising anthropogenic pollution [23], industrial gas leaks [24], disaster warning, and agricultural monitoring [25]. In the last few decades, laser-based sensing positioned itself as a compelling solution for gas detection. The *in-situ*, online, non-intrusive, fast-response, and portable characteristics of laser-based sensors make them highly applicable over a wide range of conditions and scenarios [23]. Laser-based remote sensing from HAPS offers significant potential for enhancing environmental monitoring, disaster warning, and pollution control [10,26]. By employing technologies such as LIDAR (Light Detection and Ranging) and differential absorption LIDAR (DIAL), HAPS can remotely sense various environmental parameters with high precision and in near real-time [27–32]. HAPS-based remote sensing offers several advantages over terrestrial (ground-based) sensing, particularly in terms of coverage, flexibility, and operational efficiency. One of the key advantages is the wider coverage area that HAPS can achieve due to their elevated position in the stratosphere. This allows a single HAPS to monitor large regions continuously, providing a broader perspective compared to ground-based sensors, which are limited by the Earth's curvature and obstacles like mountains or buildings [7,10]. Additionally, HAPS can access remote or inaccessible areas, such as oceans, forests, or disaster-stricken regions, where ground-based sensing infrastructure may be difficult or impossible to deploy [12,33]. Unlike satellites, which depend on fixed orbits, HAPS can hold position or adjust movement as required, offering continuous monitoring and improved temporal resolution.

HAPS not only surpass satellites in spatial resolution but also rival or outperform ground-based systems across a wide range of applications. Since HAPS operate at lower altitudes than satellites, they can achieve fine-grain detail in imagery and sensing, making them ideal for applications like environmental monitoring, disaster response, or agriculture. Furthermore, HAPS can be deployed more flexibly and quickly than ground-based systems, which may require extensive setup or infrastructure [8]. This rapid deployability is crucial for time-sensitive applications, such as disaster warning or emergency response, where immediate situational awareness is needed. This capability is especially valuable for monitoring greenhouse gas emissions, particulate matter, and other pollutants that contribute to global warming and air quality degradation. By detecting changes in these pollutants, HAPS can help regulators and policymakers make informed decisions to control pollution and limit the impact of climate change. Additionally, the ability of HAPS to continuously monitor vast geographical areas from the stratosphere makes it an excellent tool for tracking environmental trends over time [34].

In terms of disaster warning, laser-based sensing from HAPS can play a crucial role in the early detection of volcanic emissions, forest fires, and other atmospheric anomalies that may signal impending natural disasters. For example, DIAL systems can be used to detect the concentration of carbon dioxide ($CO_2$) and other gases emitted by volcanic eruptions, allowing for early warnings and evacuation planning [35]. Similarly, laser-based sensing can be employed to detect methane leaks from industrial sites, pipelines, or natural sources, thereby contributing to both environmental protection and disaster risk reduction [22]. With HAPS providing continuous coverage over sensitive regions, authorities can respond more swiftly to environmental threats, potentially saving lives and mitigating damage to ecosystems.

One of the most significant challenges in deploying HAPS for extended periods is minimizing the weight and size of the payload. The importance of this is paramount for several reasons. First, a lighter payload enables the platform to achieve greater altitudes, which enhances the coverage area for both sensing and communication [9]. Second, reducing the payload weight allows HAPS to stay aloft for longer periods, thus increasing its endurance of its monitoring and communication missions. This extended endurance is especially important for long-term environmental monitoring projects, such as tracking pollution levels or monitoring global warming indicators over weeks or months. Moreover, minimizing the payload reduces the operational costs of the platform by requiring less energy for sustained flight, which is particularly beneficial for solar-powered HAPS that rely on efficient energy use to remain in the stratosphere [36]. Traditionally, separate payloads were needed for communication and sensing. However, since both functions rely on an optical link, a single dual-function system can be realized using one optical channel [37]. Integrated sensing and communication (ISAC) is expected to reduce payload size, hardware requirements, power consumption, and overall weight [38]. By combining laser-based DIAL with optical communication using the same laser line (in THz and free-space optical (FSO) ranges, e.g., 780–850 nm and 1520–1600 nm), HAPS can transmit real-time data about air quality, greenhouse gas concentrations, and other environmental factors, offering a multi-functional platform that can serve both communication and remote sensing needs simultaneously [39].

This tutorial-cum-survey examines and reviews recent advances in HAPS-based optical communications while offering a tutorial perspective on their integration with atmospheric remote sensing through shared optical links. We evaluate various link configurations, system architectures, and sensing modalities, emphasizing their operational feasibility, spectral sensitivity, and atmospheric compatibility at altitudes around 20 km. We also examine the relevance of the ISAC paradigm in shaping future HAPS missions that combine environmental monitoring with broadband connectivity. Through detailed spectral simulations and detection limit analysis, we identify promising wavelength regions for DIAL to target key atmospheric species. A scientometric analysis is conducted to contextualize current research trends and highlight the growing interest in HAPS-based sensing. Our findings are synthesized into comparative tables that highlight the trade-offs, platform-specific considerations, and potential of HAPS-based systems for high-resolution, wide-area environmental monitoring.

### A. Existing Survey/Tutorial Articles

Recent literature on HAPS spans a wide range of perspectives, reflecting the technology's evolution from aeronautical studies to next-generation communication and sensing architectures. Previous work by d'Oliveira et al. [40] reviewed HAPS developments from an aerospace-engineering standpoint, while Qiu et al. [41] envisioned a software-defined, cross-layer network combining high- and low-altitude platforms. Later surveys such as Arum et al. [42] and Karabulut Kurt et al. [43] focused on communication-centric advancements, addressing network optimization, resource allocation, and topology management. More conceptual studies, including Mershad et al. [44], explored novel paradigms like HAPS–cloud integration, whereas recent works—Lou et al. [8], Abbasi et al. [7], and Belmekki et al. [10]—expanded toward 6G, ISAC, and smart-community applications. Distinctly, the present tutorial-cum-survey integrates optical communication and atmospheric sensing within a single framework, emphasizing DIAL and wavelength co-use for dual-purpose operation. A comprehensive comparison of these studies across multiple criteria is summarized in Table I.

TABLE I
SUMMARY AND COMPARISON OF HAPS RELATED SURVEY PAPERS.

| Survey | Focus | HAPS overview, types and config. | Earth observations from HAPS | ISAC & wavelength analysis | Challenges | Scientometrics & future trends |
|---|---|---|---|---|---|---|
| d'Oliveira et al. (2016) [40] | Reviews then contemporary HAPS technology through an aerospace-engineering lens, rather than communication-focused one. | ⊕ | ⊖ | ⊖ | ⊕ | ⊕ |
| Qiu et al. (2019) [41] | Proposes a software-defined, cross-layer integrated network architecture uniting high- and low-altitude platforms with terrestrial cellular systems. The study further outlines potential use cases, identifies their challenges, and demonstrates feasibility through a proof-of-concept case study. | ⊕ | ⊖ | ⊖ | ⊕ | ⊖ |
| Arum et al. (2020) [42] | Provides an overview of HAPS potential for delivering services in remote regions and examines enabling methods for intelligent radio resource allocation, adaptive topology management, and extended coverage. | ⊕ | ⊖ | ⊖ | ⊕ | ⊖ |
| Karabulut Kurt et al. (2021) [43] | Presents an extensive survey of contemporary HAPS network research, outlining future opportunities and associated challenges. It further discusses key aspects of HAPS design, resource allocation, and topology management principles. | ⊕ | ⊖ | ⊖ | ⊕ | ⊖ |
| Mershad et al. (2021) [44] | Provides a high-level conceptual overview of an emerging technology rather than a survey of existing literature. Introduces the interplay between HAPS and cloud computing, | ⊕ | ⊖ | ⊖ | ⊕ | ⊖ |

| Survey | Focus | HAPS overview, types and config. | Earth observations from HAPS | ISAC & wavelength analysis | Challenges | Scientometrics & future trends |
|---|---|---|---|---|---|---|
| | highlighting applications enabled by this integration and discussing their associated methods and challenges. The study represents an initial step toward developing innovative HAPS–cloud frameworks for next-generation (6G) flying networks. | | | | | |
| Lou et al. (2023) [8] | Analyzes HAPS as a key component of non-terrestrial networks, emphasizing its role in linking satellites, aerial systems, and ground infrastructure. Presents prospective 6G architectures, discusses enabling optical and mmWave technologies, and outlines key performance metrics. The study highlights HAPS' potential for extended coverage and low-latency service delivery while identifying challenges in energy use, coordination, and channel modeling. | ⊖ | ⊖ | ⊖ | ⊕ | ⊖ |
| Abbasi et al. (2024) [7] | Reviews potential HAPS use cases within 6G, positioning them as platforms for communication, computing, and sensing integration. The paper discusses ISAC concepts, system design issues, and atmospheric constraints while mapping HAPS capabilities to emerging 6G demands. It concludes by outlining challenges in mobility management, beamforming, and multi-layer network interoperability. | ⊖ | ⊖ | ⊖ | ⊕ | ⊖ |
| Belmekki et al. (2024) [10] | Examines HAPS-assisted aerial networks for people-centered smart communities and broadband access. Describes HAPS architectures, operational frameworks, and a proof-of-concept 5G trial. The paper underscores HAPS' role in extending connectivity, improving resilience, and bridging digital divides while noting open issues in energy efficiency and standardization. | ⊕ | ⊕ | ⊖ | ⊖ | ⊖ |
| This tutorial/survey | Presents a tutorial-cum-survey on integrating HAPS-based optical communication with atmospheric sensing. The paper reviews HAPS architectures, link geometries, and sensing modalities, emphasizing the potential of Differential Absorption Lidar (DIAL) for trace-gas detection. It provides a spectral roadmap for dual-use telecom wavelengths, discusses payload design and operational trade-offs, and outlines future research directions. The work uniquely bridges broadband connectivity and environmental monitoring within a unified stratospheric platform concept. | ⊕ | ⊕ | ⊕ | ⊕ | ⊕ |

*Notation:⊕ means fully covered; ⊖ means partially covered.*

*B. Contributions and Structure*

Despite the accelerating progress in both stratospheric communications and atmospheric sensing, these domains have largely evolved along parallel trajectories, with minimal hardware or operational convergence. Traditional remote-sensing missions rely on dedicated satellite or airborne instruments that deliver high scientific fidelity but suffer from long revisit times, limited spatial resolution, and high deployment costs. Conversely, free-space optical (FSO) communication systems on HAPS have achieved impressive data-rate milestones but are typically optimized for unidirectional information transfer, neglecting the potential of their optical payloads as scientific sensors. This separation of functions results in redundant hardware, excess payload mass, and sub-optimal energy utilization—critical limitations for long-endurance stratospheric platforms.

The motivation for this work stems from the need to unify these two optical disciplines—communication and sensing—into a single, co-optimized framework that maximizes the scientific and operational value of each photon transmitted from the stratosphere. High-Altitude Platform Stations (HAPS) provide an unprecedented vantage point for achieving this synthesis. Operating between 17 and 22 km altitude, above most weather and turbulence yet below orbital constraints, they can sustain persistent line-of-sight connectivity while performing continuous environmental observation. The emergence of Integrated Sensing and Communication (ISAC) architectures and the maturation of telecom-grade photonics create the technological opportunity to realize such dual-use optical payloads.

Accordingly, the objectives and contributions of this paper are fourfold:

1) *Comprehensive synthesis*: We provide a tutorial-style review of HAPS technologies, spanning platform types, link geometries, and atmospheric transmission conditions, to establish a physical and operational basis for optical sensing and communication in the stratosphere.

2) *Methodological integration*: We examine the compatibility of laser-based remote-sensing techniques—especially Differential Absorption Lidar (DIAL)—with existing FSO communication terminals, identifying opportunities for shared optics, wavelength-division multiplexing, and unified data handling.

3) *Spectral roadmap and feasibility analysis*: Through detailed HITRAN-based simulations, we identify absorption micro-windows in the telecom C-band (1.53–1.57 μm) suitable for trace-gas retrieval while preserving high-capacity optical downlinks, quantifying expected detection limits and sensitivities for key species ($CO_2$, $CH_4$, $N_2O$, $O_3$, $H_2S$).

4) *Research context and future outlook*: We conduct a scientometric analysis of two decades (2005–2025) of HAPS-related literature to reveal disciplinary trends, institutional leadership, and emerging cross-domain collaborations that underpin the convergence of communication and environmental monitoring.

The remainder of this paper is organized as follows: Section II outlines the main types and configurations of HAPS platforms and configurations. Section III contrasts stratospheric platforms with terrestrial and orbital architectures in terms of coverage, latency, and operational flexibility. Section IV examines optical link geometries and atmospheric considerations, including nadir, slant, and cross-beam configurations, as well as turbulence and pointing stability. Section V reviews major atmospheric remote-sensing techniques and evaluates their compatibility with HAPS, emphasizing the suitability of Differential Absorption Lidar (DIAL). Section VI presents the theoretical fundamentals of DIAL, deriving its key retrieval equations and sensitivity functions. Section VII explores the role of HAPS in DIAL-based environmental monitoring and outlines potential application domains. Section VIII introduces the concept of ISAC on HAPS, describing system architectures, spectral compatibility, and operational co-design. Section IX identifies optimal spectral bands and analyzes the overlap between sensing and telecom windows from the ultraviolet to the mid-infrared. Section X discusses optical payload constraints—including size, weight, power, thermal management, and reliability considerations—while Section XI presents a scientometric analysis of the HAPS research landscape, highlights future trends, and concludes with perspectives on next-generation ISAC-enabled HAPS observatories. Section XII concludes the paper with the main findings and recommendations.

## II. HAPS TYPES AND CONFIGURATIONS

HAPS are stratospheric unmanned platforms (typically at ~20 km altitude) that serve as "pseudo-satellites" for communications or observation [45]. HAPS platforms generally fall into two broad categories: (a) Fixed-wing, solar-powered aircraft (including both unmanned and manned versions) and (b) Lighter-than-air vehicles (such as high-altitude airships or long-duration balloons). Fixed-wing HAPS (e.g., Airbus *Zephyr*) usually circle in a tight flight pattern to remain over a target area, relying on solar panels and batteries for continuous operation. Lighter-than-air HAPS (e.g., *Sceye*, and *Thales Alenia Stratobus* concepts [46,47]) are designed to hover or drift only minimally, using buoyant lift and sometimes gentle propulsion to station-keep in the stratosphere. These platforms can stay aloft for weeks to months, offering a stable "eye in the sky" far above weather systems. Modern HAPS designs leverage advanced lightweight materials, high-efficiency solar cells, and energy-dense batteries to achieve long endurance (on the order of 30+ days) [48]. For instance, recent HAPS projects span from fixed-wing solar drones (*Zephyr, BAE PHASA-35, SoftBank/HAPSMobile Sunglider/Hawk30*) to balloon/airship systems (*Loon superpressure balloons, Stratobus*) [5]. HAPS can operate either as standalone units or in constellations, and can host multi-purpose payloads. In a standalone "tower in the sky" configuration, a single HAPS could provide wide-area coverage (radius of tens of kilometers) for communications or sensing. Alternatively, multiple HAPS can be networked to cover larger regions or to hand off missions, analogous to how multiple cell towers or satellites operate in tandem [45]. The flexibility in platform type and deployment configuration allows HAPS to be tailored to specific mission needs, for example, a fixed-wing UAV HAPS might be preferred for rapid redeployment and maneuverability, whereas an airship HAPS could be advantageous for continuous long-term station-keeping over a fixed spot. Notably, HAPS fill a niche between satellites and conventional aircraft: they have much lower latency and path loss than low-Earth orbit (LEO) satellites (due to the closer range), and can carry larger payloads with longer observation time than small UAV drones [5]. This unique positioning makes HAPS an attractive platform for applications like persistent environmental monitoring and broadband connectivity.

A clear understanding of HAPS platform types and their capabilities is essential before turning to link and payload considerations, as integration strategies depend heavily on platform characteristics. At stratospheric altitudes, HAPS are realized in several forms, each with specific advantages.

*A. Stratospheric Balloons*

These range from zero-pressure balloons (e.g., *Aerostar,* which vent gas to stabilize altitude) to super-pressure balloons (sealed to maintain lift). Balloons are relatively low-cost and lightweight, which simplifies launch and deployment. However, by nature they drift with winds and cannot be actively station-kept over a precise location (at best, limited steering is achieved by changing altitude to catch different wind layers) [49]. They also have limited power and payload capacity, often carrying only a few kilograms with power on the order of tens of watts (solar+battery systems) for long-duration flights. Despite these constraints, balloons like World View's *Stratollite* have demonstrated multi-week station-keeping within a radius of tens of kilometers, sufficient for many monitoring tasks [50]. Balloons are well-suited for applications where broad coverage is needed more than precise positioning, and for initial tech demonstrations (they have lifted optical communication experiments such as *STROPEX*, discussed later).

*B. Fixed-Wing High-Altitude UAVs*

Solar-powered unmanned airplanes (e.g., Airbus *Zephyr*) can actively hold position over a target area and carry larger payloads than balloons. Modern designs can lift payloads in the mid tens of kilograms and supply a few hundred watts of electrical power continuously to onboard systems [49]. They achieve this via efficient aerodynamic design and solar arrays charging batteries for night use. These UAVs can remain aloft for months in theory, though in practice current endurance records are on the order of weeks. The advantage of fixed-wing HAPS is the ability to precisely orbit over a specific point (or systematically patrol a region) and to host more complex payload suites (for example, a combination of optical communication terminal, LIDAR, and cameras). The trade-off is that such aircraft are highly optimized for low weight, typically made of ultralight composites, so every additional kilogram of payload or watt of power must be carefully accounted for in the design, which motivates ISAC.

*C. High-Altitude Airships*

Stratospheric airships (lifting by helium like balloons but with powered propulsion and rudders) promise the best of both worlds: station-keeping capability with very large payloads. Next-generation designs (e.g., projects by *Sceye, Thales*, etc. [46,47]) envision hundreds of kilograms payload and power levels of multiple kilowatts (via large solar arrays) [49]. An airship's large surface area supports more solar cells, and its buoyancy can carry heavier optical instruments (like sizable telescopes or high-power laser systems that a smaller UAV or balloon could not lift). They also potentially offer longer mission durations (months to a year). The complexity of airships is higher; they are large (tens of meters long), require advanced materials to handle stratospheric conditions, and present a bigger radar/visual. While few have been fully tested operationally, airships could in the future provide a stable platform for ISAC, effectively a persistently hovering stratospheric base station.

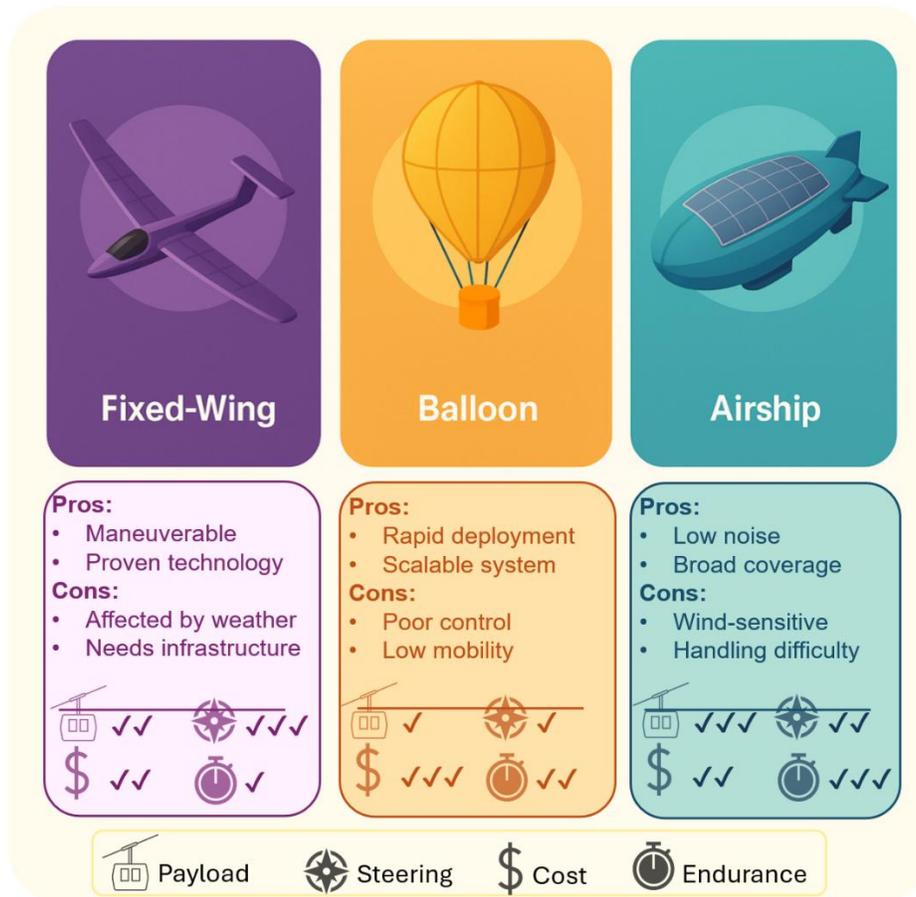

**Fig. 1.** Comparison of HAPS types (Fixed-Wing, Balloon, and Airship), highlighting their operational advantages, limitations, and performance across four key metrics: Payload capacity, steering capability, cost, and endurance.

All these platforms operate in the extreme environment of ~20 km altitude where air pressure is only about 7% of sea-level and temperatures can be around -50°C. These conditions mean reduced aerodynamic lift, but also negligible weather interference (above storms and clouds). HAPS generally fly above regulated airspace (in many jurisdictions, air traffic control ends at ~18 km/60,000 ft [49]), simplifying flight logistics and avoiding conflicts with commercial aircraft. This "near space" regime between airplanes and satellites is now increasingly seen as valuable real estate for routine operations. The choice of HAPS will influence how an optical communication and sensing payload is designed and integrated. Fig. 1 summarizes the three main configurations of HAPS with their unique characteristics and advantages.

## III. COMPARATIVE BENEFITS: HAPS VS. GROUND AND SATELLITE SYSTEMS

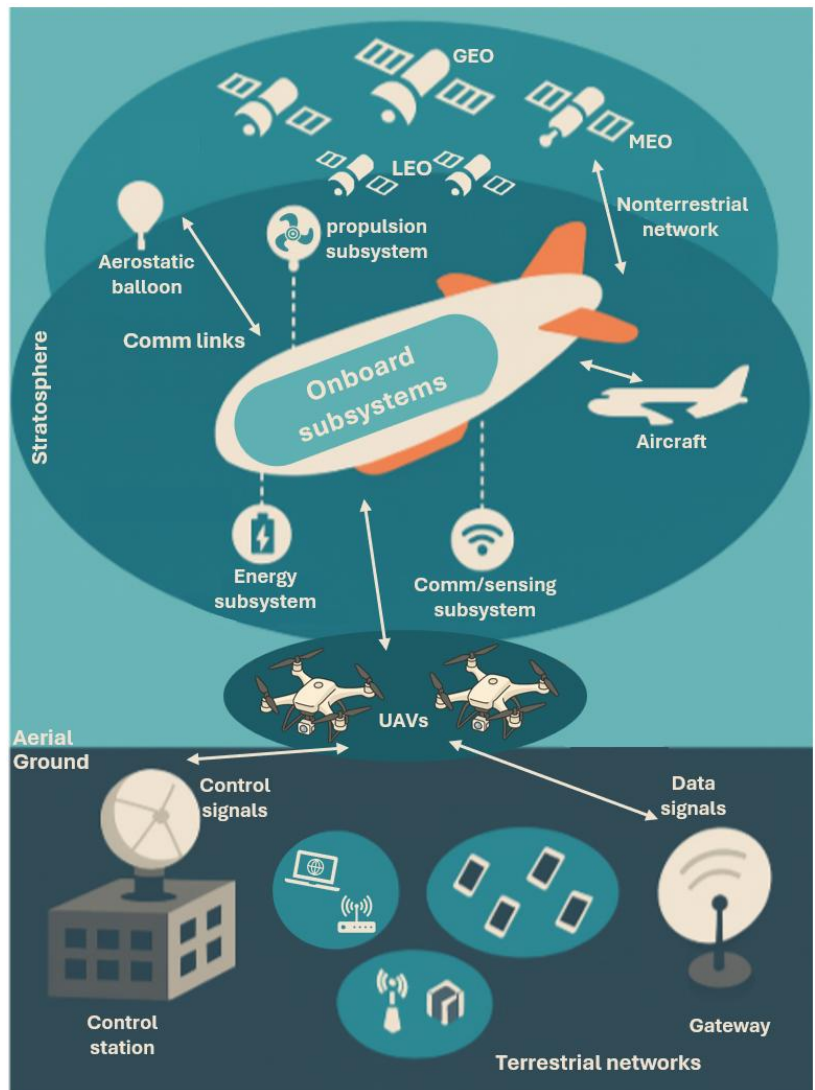

**Fig. 2.** Three-tier non-terrestrial network (NTN). satellites in geostationary (GEO), medium-earth (MEO) and low-earth (LEO) orbits link with haps.

HAPS-based optical communication and sensing can be understood as part of a layered observation and communication infrastructure. They complement and fill gaps between ground-based and satellite-based approaches. Fig. 2 demonstrates this concept. The stratospheric HAPS layer forms a pivotal "middle mile" between terrestrial sites and orbital assets. By station-keeping at 18–25 km, HAPS platforms maintain quasi-geostationary footprints that overcome terrain shadowing and atmospheric turbulence that often degrade long ground links, while avoiding the large path losses and latency penalties inherent to geostationary/medium-Earth orbits (GEO/MEO). Their onboard optical payloads can down-link high-throughput data to gateway stations, hand off traffic laterally to neighboring HAPS or UAV swarms for local distribution, and up-link aggregated user traffic or Earth-observation data to LEO relays. In this way, the layered stack (satellites for global reach, HAPS for regional agility, UAVs for last-kilometre flexibility, and terrestrial networks for end-user access) creates a resilient, multi-hop optical backbone that closes spatial and temporal coverage gaps left by any single tier alone. In the following, we compare HAPS with ground systems and satellites to highlight the unique value of HAPS.

### A. HAPS vs Ground Systems

Ground-based optical communication links (like point-to-point free-space laser links or ground-to-satellite uplinks) suffer from the thick atmosphere and curvature of the Earth limiting line-of-sight. Terrain, buildings, and Earth's curvature allow only a few tens of kilometers of range near ground before the beam is blocked or attenuated. In contrast, a single HAPS at 20 km altitude has a wide coverage area; it can see the ground up to ~500 km away. This means one HAPS could replace a chain of multiple ground towers in communications coverage. Additionally, ground optical links are severely affected by weather (rain, fog, haze can all disrupt lasers). A HAPS above the weather avoids local fog or mist; it only contends with those in the terminal area of the ground station. For sensing, ground instruments like LiDAR ceilometers or air quality laser sensors are fixed in location and typically provide data for a single column of air or along a single horizontal path. They cannot readily map large areas. A HAPS carrying a laser sensor can overfly and map distributions over hundreds of square kilometers, offering a synoptic view impossible from one ground site. Moreover, positioned above the boundary layer, the HAPS sensor avoids turbulence and pollution, enabling top-down measurements that can improve accuracy for certain observations. That said, ground systems are indispensable for calibration and local detail; a network of ground gas sensors could calibrate the HAPS's remote sensing algorithm, and ground fiber networks ultimately distribute the data that HAPS downlinks provide. So rather than redundant, they are complementary: HAPS extends the reach of ground systems to the stratosphere, providing mobility and high vantage, while ground assets provide persistent presence when the HAPS has moved on or for continual monitoring at fixed stations.

### A. HAPS vs Satellites

Satellites cover broad areas and can reach truly global scale, but they have well-known limitations: orbital spacecraft either move quickly relative to the ground (LEO satellites orbit Earth in ~90 minutes, so they are over a given area for only a few minutes), or if geostationary, they are 36,000 km away which results in long signal travel times and poor spatial resolution for sensing. HAPS, by flying 1–2 orders of magnitude closer to Earth, can achieve much higher data rates and resolution and with much lower latency. For instance, the round-trip latency to a geosatellite is about 240 ms, whereas to a HAPS it is only ~0.13 ms, thus enabling real-time control and low-latency communications [51]. In terms of resolution, a hyperspectral imager on a HAPS might have a ground sampling distance of a few meters [52], resolving individual buildings or plumes, while a satellite sensor might be limited to tens or hundreds of meters per pixel. HAPS can thus perform "satellite-like" functions with drone-like detail. Another key difference is persistence: A single HAPS can dwell over one area continuously (quasi-geostationary with respect to the ground) [51], whereas a single LEO satellite provides a brief snapshot and then the area waits for the next orbit. Achieving continuous coverage by satellite requires a large constellation (dozens of satellites in coordinated orbits), which is costly. A HAPS, essentially a reusable pseudo-satellite, can be deployed to hotspot areas on demand, for example, send a HAPS to hover over a hurricane impact zone for a week, rather than needing a pre-existing satellite in the right orbit. In terms of cost and deployment speed, HAPS have an advantage: They can be launched and brought back for maintenance relatively quickly and at far lower cost than building and launching a satellite [51]. This makes them ideal for technology testing (trying new sensors or communication devices without the high stakes of a satellite launch) and for surge capacity (e.g., deploying extra capacity during an event or conflict).

Satellites complement HAPS by covering areas the latter cannot, such as expansive oceans and high-latitude regions where solar HAPS are ineffective during winter. Unlike HAPS, satellites are not influenced by local weather at the platform, though cloud cover remains a limiting factor for ground observation. Satellites also handle very large-scale data like global greenhouse gas mapping, whereas HAPS might focus on regional scales. The optimal future scenario likely involves a synergy as shown in Fig. 2: Satellites provide a broad overview and long-term datasets, while HAPS fill in with high-resolution, real-time data in targeted areas, and ground networks provide ground-truth and high-density local data. For communications, satellites (especially proposed mega-constellations of laser-linked LEO satellites) aim to deliver internet globally, but even they face challenges like latency and capacity in dense areas; HAPS could serve as a middle layer, receiving high-speed feeds from satellites and then using their proximity to users to distribute bandwidth where needed (a concept sometimes called "layered space-air-ground networks" [34,53]). HAPS laser links to satellites have already been considered to offload data quickly from Earth observation satellites (acting as intermediate relay). In defense scenarios, HAPS can cover the gap when satellites are not available (due to orbit or if they are jammed/destroyed) by providing theater-level communications and intelligence, surveillance, and reconnaissance (ISR).

HAPS offer a middle ground that combines some strengths of ground systems (flexibility, upgradeability, proximity resulting in high data throughput) with some strengths of satellites (wide coverage, high vantage point). When equipped with both optical communications and sensors, HAPS become powerful "eye in the sky" platforms that not only observe but also instantly communicate, something that neither ground sensors nor satellites alone can do as effectively. This comparative edge is driving interest in HAPS as key assets in future networks and observing systems. Table II summarizes the comparison between HAPS and ground systems and satellites.

TABLE II
COMPARATIVE FRAMEWORK OUTLINING KEY ASPECTS ACROSS THREE PLATFORM CATEGORIES: GROUND SYSTEMS, HAPS, AND SATELLITES.

| Aspect | Ground systems | HAPS | Satellites |
|---|---|---|---|
| **Line-of-sight / coverage footprint** | Limited to ≈10–50 km by terrain & Earth curvature; many towers needed for wide areas | One HAPS at ≈20 km sees a disk ≈400–600 km in radius, replacing long chains of ground relays | GEO or wide swaths (LEO) but with large footprints that dilute resolution |
| **Atmospheric effects on optical links** | Full atmospheric column; rain, fog, haze heavily attenuate lasers | Above most weather; only ground-terminal segment affected | Links space–ground traverse full atmosphere; downlinks blocked by clouds for optical communications /sensing |
| **Latency (two-way)** | Fiber: sub-ms; ground FSO ≈ 0.1–1 ms | ≈ 0.1 ms (20 km altitude); real-time control feasible | GEO ≈ 240 ms; LEO ≈ 20–40 ms |

| Aspect | Ground systems | HAPS | Satellites |
|---|---|---|---|
| **Spatial resolution for imaging/sensing** | Very high at point of measurement but geographically sparse | Meter-scale ground sampling distance possible over hundreds km²; can dwell for continuous mapping | Tens–hundreds m (LEO) or km-scale (GEO); revisit or resolution limited by orbit & altitude |
| **Persistence over a target** | Fixed sensors provide 24/7 data at a single site/path | Quasi-stationary over region for days–weeks; redeployable as needed | LEO: minutes per pass, long gaps unless large constellation; GEO: continuous but coarse |
| **Weather / cloud avoidance for remote sensing** | Instruments look through boundary-layer haze & turbulence | Can fly under cloud decks or above them, choosing best vantage | Fully above weather, but clouds block downward view; no option to get closer |
| **Deployment speed & upgradeability** | Instant hardware access; inexpensive | Launch/land from airfields; sensors/payloads swappable within days | Years of development & launch; no hardware upgrades after launch |
| **Cost (per platform)** | Lowest (existing towers, fiber); scaling requires many sites | Moderate: reusable airship/solar UAV << satellite bus & launch | High: launch + bus; even higher for multi-satellite constellations |
| **Ideal roles** | Local backhaul, calibration, dense in-situ sensing | Regional high-resolution sensing, low-latency communications relay, surge capacity, tech demos | Global coverage, long-term climate data, oceanic/remote regions |

## IV. Link Geometry and Atmospheric Considerations

Optical link geometry from a HAPS encompasses three main scenarios: (1) Downlink/Uplink (Nadir), i.e., near-vertical beams between the HAPS and the ground directly below, (2) Slant Paths: diagonal links between the HAPS and a ground station or target at some horizontal distance, and (3) Cross-Platform Links (Horizontal): nearly horizontal beams between two HAPS or between HAPS and satellites above. Each has different implications for range and atmospheric penetration as illustrated in Fig. 3.

### A. Nadir Links

A HAPS directly above a ground terminal or target has the shortest path (roughly the altitude, 20 km, or slightly more if at an angle). This minimizes atmospheric traversal; the beam passes through the lower atmosphere nearly perpendicularly, reducing the path length through dense air. For communications, a nadir downlink can deliver high power density to a receiver on the ground, supporting high data rates. For gas sensing, a nadir-looking LIDAR can profile the vertical column below the HAPS (e.g., measuring gas concentration as a function of altitude by timing the LIDAR return). Many HAPS missions will involve a near-nadir geometry, such as linking to a dedicated optical ground station within the HAPS's overhead footprint or scanning the local area beneath for emissions.

### B. Slant Links

If the ground station or area of interest is not directly under the HAPS, the optical beam will take a slant route through the atmosphere. For example, a HAPS might need to connect to a ground hub that is 100 km away horizontally; the beam then goes through 20 km of vertical and 100 km of horizontal distance in air. Such a path traverses a longer stretch of troposphere, increasing attenuation from scattering and absorption. For communications, this means more loss and potentially more turbulence-induced fading; for sensing, the beam may integrate gas absorption over a longer path that complicates pinpointing location. HAPS designers mitigate this by either keeping links as vertical as possible or using multiple HAPS hops to cover distance. For instance, rather than a single 300 km slant link from HAPS to ground, one might send data HAPS-to-HAPS horizontally at altitude, then downlink when nearly overhead the destination. Indeed, at 12.5 km altitude a line-of-sight reaches ~400 km, and at 20 km it exceeds 500 km, so a network of HAPS spaced ~200–300 km apart could leapfrog in the stratosphere with each downlink occurring under favorable (near-vertical) conditions [54]. This multiple HAPS concept motivates the last link configuration, horizontal links.

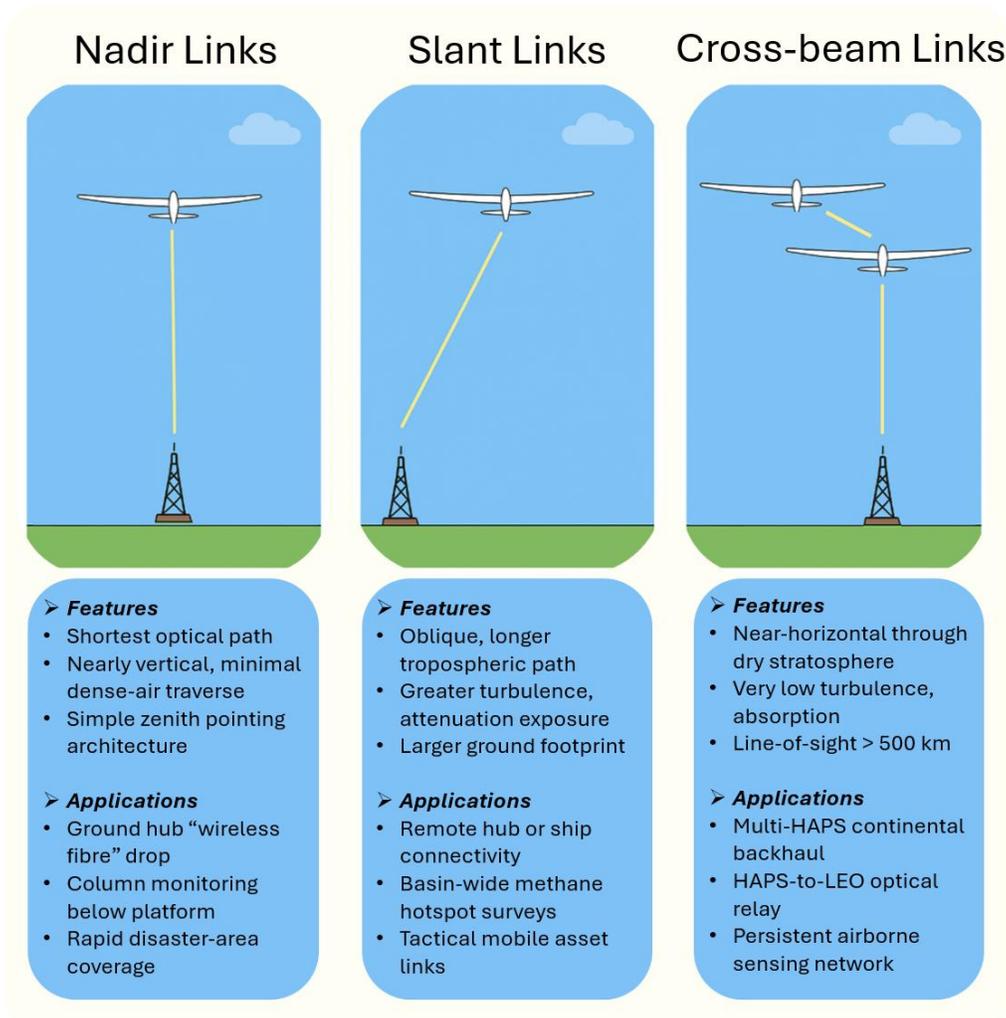

**Fig. 3.** Three types of optical links used in HAPS: Nadir Links, Slant Links, and Cross-beam Links.

*C.  Horizontal (Cross-Beam) Links*

A unique advantage of stratospheric platforms is the ability to link to peers or even to satellites through very thin atmosphere. Two HAPS at 20 km can have a direct optical line between them largely above clouds. The intervening air at ~20 km is extremely dry (above most water vapor) and thin, so absorption and turbulence are minimal. This can enable multi-HAPS optical networks carrying broadband data over continental scales. In fact, one HAPS-to-HAPS optical link was projected to achieve over 100 km range in tests, with multi-Gbps capacity, essentially creating an airborne fiber-optic network [55,56]. Similarly, a HAPS could use optical links to LEO satellites passing above; being so high up, the HAPS can communicate to space with very little atmospheric interference (a big advantage for feeding satellite internet systems). These cross-links benefit from the stratospheric calm; turbulence strength is orders of magnitude lower at 20 km than near the ground, so optical beams wander and scintillate much less on these paths. Overall, horizontal links at altitude are the most reliable optical paths available in the atmosphere, often limited only by geometric line-of-sight and curvature of the Earth.

*D.  Atmospheric Transmission Characteristics*

HAPS altitudes are generally favorable for optical propagation, but with important caveats. Being above ~95% of the atmospheric mass, a HAPS does avoid most of the lower-atmosphere absorption and weather effects. Notably, at ~20 km a platform is above the cloud deck; in the stratosphere, clouds are rare (only occasionally nacreous or noctilucent clouds at those heights), and above the bulk of water vapor and aerosols. This means that for uplink/downlink communication, the majority of attenuation and turbulence occurs in the lowest few kilometers of the beam path (near the ground). A HAPS optical link thus experiences significantly less fading than a ground-level horizontal link of the same distance. It was demonstrated, for instance, that a stratospheric optical downlink can maintain a high data rate of 1.25 Gbps to a ground station, with the main impairments being near-ground turbulence cells and not the stratosphere [57]. In that 22-km balloon experiment (the *STROPEX* campaign), scientists measured turbulence profiles and found the stratospheric portion contributed very little distortion compared to the lower tropospheric portion [57]. The thin air also reduces Rayleigh scattering for optical wavelengths, benefiting visibility (which can exceed 300 km at altitude on clear days).

However, when looking downward for gas sensing, a HAPS's laser still has to traverse the thicker atmosphere below to reach the surface or the gas plumes of interest. Absorption by background gases (e.g., $CO_2$ and $H_2O$) along that path must be considered. For instance, any open-path DIAL from HAPS will need to account for intervening layers. A key benefit of HAPS is the ability to position the sensor for optimal sensitivity, whether observing from above a pollution layer or probing horizontally through it. The absence of clouds is crucial; clouds are essentially opaque at most laser wavelengths, so HAPS sensing operations would be scheduled during clear-sky conditions over the target or use alternative viewpoints (multiple HAPS or angles) if partial cloud cover exists [57]. Strategies like ground station diversity (using multiple ground receivers for communication) have been proposed to overcome cloud outages [57]; similarly, multi-angle sensing or waiting for cloud gaps are solutions for gas observation.

*E.  Stabilization and Tracking Systems*

HAPS optical payloads employ gimbaled optical mounts and fast steering mirrors that can adjust the beam direction in real time. Typically, a pointing, acquisition, and tracking (PAT) system is used: for instance, a HAPS downlink might send a weak beacon laser or use an LED guide star that a ground station can track, and vice versa, enabling closed-loop feedback to constantly correct the pointing [57]. In the *STROPEX* demonstration, a transportable optical ground station tracked the incoming balloon laser and instrumentation measured the beam wander caused by turbulence [57]. Based on such feedback, actuators on the transmitter can nudge the beam to compensate for drifts. Modern optical terminals often include inertial measurement units (IMUs) and star trackers to know the orientation of the platform, along with feedback from the receiver (via uplink beacon or retro-reflector) to fine-tune alignment. This multi-tiered approach is essential because a HAPS can have slow drift (e.g., a balloon moving at a few m/s) superimposed with higher-frequency jitter (e.g., a UAV's slight wing flutter or an airship's vibration).

*F.  Platform Motion and Control*

Different HAPS have different stability characteristics. A balloon might rotate slowly as it drifts, potentially twisting cables or altering the orientation of fixed-mounted optics. Thus, balloon-borne optical systems often need a motorized rotation stage to counteract balloon rotation and keep pointing toward the ground station. Fixed-wing UAVs, on the other hand, may fly in circles to maintain a relative position. This means the optical payload might continuously need to pan to keep line-of-sight to a fixed ground target. A carefully designed flight pattern (such as a "lazy eight" or continuous orbit) combined with a gimbaled turret can achieve a quasi-steady link. Airships can hold a fixed orientation more easily by adjusting rudders and propellers, making pointing somewhat easier, though they may still oscillate around a tether point. In all cases, autopilot and stabilization algorithms are critical; for example, holding a UAV within ±0.5° of a desired attitude, so the fine steering mirror only cleans up the remainder.

*G.  Pointing Accuracy Requirements*

The required accuracy depends on beam width and receiver aperture. For a communication link using a 10 cm transmitting telescope and a diffraction-limited laser at 1550 nm, divergence might be on the order of tens of μrad. To couple a significant fraction of power into a 20 cm ground telescope, one might need pointing errors $< \sim$100 μrad. This is challenging but achievable with electro-optical tracking. For sensing, if the goal is to scan a specific facility (say a methane plume over a gas well), the pointing might need to be stabilized to within a few tens of meters on the ground, equivalent to a few μrad accuracy, to repeatedly sample the same spot. Any platform motion that deviates the aim beyond this must be corrected promptly.

*H. Turbulence and Beam Wander*

Atmospheric turbulence, especially near the ground, can also deflect and blur the beam, effectively adding an apparent pointing jitter. Techniques like adaptive optics (deformable mirrors adjusting beam phase) have been used in ground astronomy and could be applied in HAPS communications to counteract this [36]. However, the complexity for a HAPS terminal may be high, so more commonly, system designers enlarge the beam a bit (to reduce sensitivity to small wandering) at the cost of some power. In *STROPEX*, for example, the downlink used a 20 cm aperture and likely a divergence that ensured the beam footprint a little larger than the ground receiver, so minor pointing errors or turbulence shifts did not break the link [57].

*I. Vibration Isolation*

Mechanical vibrations from HAPS (engines and moving control surfaces) can be a concern for optics. For instance, a propeller's oscillation might introduce a high-frequency buzz. Good practice involves isolating the optical bench with dampers and perhaps active piezoelectric stabilizers. Some high-altitude UAVs are electric (solar/battery) and have minimal vibration compared to fuel engines, which is beneficial. In the case of balloons, the gondola can swing a bit like a pendulum. Tether lines and dampers are sometimes used between the balloon and payload to reduce this motion. Additionally, software filtering in the tracking loop can distinguish between slow drift (which requires gimbal slewing) and fast jitter (which a fast-steering mirror can handle) to optimize stability.

## V. ATMOSPHERIC REMOTE SENSING TECHNOLOGIES

Modern atmospheric remote sensing encompasses a variety of techniques to observe atmospheric composition and dynamics. Key methods include passive techniques that use natural radiation (e.g., sunlight) and active techniques that use controlled illumination (lasers, radar). Below we outline several important technologies relevant to HAPS-based atmospheric sensing:

*A.  Solar Occultation*

This passive technique observes the Sun through the limb of the atmosphere during sunrise/sunset (occultation) to derive vertical profiles of atmospheric constituents. By measuring the solar spectrum after it has traversed a long path through the atmosphere, one can retrieve high-resolution vertical concentration profiles of gases and aerosols. A major advantage of solar occultation is its extremely high signal-to-noise ratio and fine vertical resolution, on the order of ~1 km or better [58]. For example, the stratospheric aerosol and gas experiment (SAGE-II) satellite instrument used solar occultation to achieve ~1 km vertical resolution profiles of ozone, $NO_2$, $H_2O$, etc. [59]. The high coherence and intensity of the Sun's light effectively provide a calibration reference, yielding very accurate measurements. However, solar occultation is limited to twice per orbit (sunrise/set) from a given platform and thus provides sparse spatial coverage. On a HAPS, solar occultation could be employed during dawn/dusk periods to obtain detailed profiles of trace gases (for instance, stratospheric ozone or aerosols) above the platform's location. The technique's reliance on direct Sun alignment means it would sample specific azimuth angles, but could complement other methods by providing benchmark profile data with minimal instrument self-calibration needed [59].

### B. Scattered Sunlight (Nadir/Limb) Measurements

Another passive approach is to observe sunlight that has been scattered by the Earth's surface or atmosphere. Instruments either look nadir (downward) to measure sunlight reflected off the ground, or limb to measure sky radiance sideways through the atmosphere. By analyzing the absorption features imprinted on this scattered light, one can infer atmospheric gas concentrations. This is the principle behind satellite UV/VIS spectrometers (e.g., GOME, OMI, TROPOMI) and ground-based differential optical absorption spectroscopy (DOAS) instruments. For example, tropospheric $NO_2$ and $SO_2$ have been measured by fitting the UV-visible absorption fingerprints in backscattered sunlight spectra [60]. Nadir-viewing remote sensing of scattered sunlight provides wide horizontal coverage and is suitable for continuous monitoring during daytime. However, the measurements yield integrated column amounts or coarse vertical information (without the sharp resolution of occultation) because the path of photons is complex and requires modeling. In practice, algorithms use radiative transfer models and air mass factors to convert the slant column absorption into vertical column densities [60]. For instance, retrieving tropospheric $NO_2$ from a nadir sensor involves accounting for how scattering geometry distributes sensitivity with altitude [60]. The advantage of the scattered-light method is that it works across broad areas in daylight, enabling global or regional maps of pollutants and greenhouse gases. From a HAPS, an instrument could observe the downward radiance or limb sky to frequently map gas distributions beneath the platform. Such a system effectively acts like a low-flying "mini-satellite", with higher spatial resolution but smaller footprint compared to orbiting sensors. One challenge is ensuring calibration and dealing with clouds or surface albedo variations that affect the measured spectra. Still, numerous studies have demonstrated that absorption spectroscopy of scattered sunlight is a robust tool for gases like $O_3$, $NO_2$, $SO_2$, HCHO, $CH_4$, etc., on scales from ground-based up to satellite [60,61].

### C. Heterodyne (Coherent) Detection

Heterodyne detection refers to techniques that mix a received optical signal with a local oscillator laser on a photodetector, effectively converting optical frequency information into a beat frequency (RF) signal. This coherent detection approach is powerful for atmospheric sensing in that it preserves phase and frequency information of the incoming light. A prime application is Doppler wind lidar: a laser pulse backscattered from aerosols or molecules is mixed with a reference laser, and the frequency shift (Doppler) is extracted from the heterodyne beat signal to deduce wind velocity along the line-of-sight [62]. Heterodyne lidar is extremely sensitive to motion, for example, wind Doppler lidars at 1.5 μm use coherent detection to measure wind speeds with precisions of <1 m/s by analyzing the intermediate frequency signal [62]. The principle is that the mixed signal produces a heterodyne current oscillating at the difference between the local oscillator and signal frequencies, which can be precisely measured in the electronic domain [62]. Besides wind measurement, heterodyne spectroscopy can achieve ultra-high spectral resolution for gas detection. For instance, infrared heterodyne spectrometers have been used to resolve narrow molecular absorption lines in the thermal IR for atmospheric analysis [63,64]. The advantage of heterodyne techniques is the high sensitivity and frequency precision, enabling detection of minute Doppler shifts or small absorbance features that direct detection might miss. On the downside, coherent detection systems are more complex and often require stable local oscillator lasers and turbulence mitigation (since atmospheric turbulence can reduce heterodyne efficiency [64]). In a HAPS context, heterodyne receivers could be used in Doppler lidars to map wind fields around the platform (useful for weather monitoring or platform navigation) or in high-resolution spectroscopy instruments to monitor trace gas abundances with fine spectral detail. The heterodyne approach essentially brings radar-like techniques into the optical regime, offering range and velocity information simultaneously for remote sensing.

### D. Airborne Onboard Sensors

In addition to "remote" sensing, HAPS (or any aircraft/drones) can carry onboard sensors or active samplers into the atmosphere. These include gas analyzers (for $O_3$, CO, $CO_2$, $CH_4$, etc.), particulate samplers, and meteorological probes. By physically being in the environment, such sensors provide ground-truth measurements of atmospheric composition at the flight altitude or through vertical profiles (if the HAPS can change altitude). For example, small UAVs have been equipped with $CO_2$ and $CH_4$ sensors and flown through plume regions to directly measure greenhouse gas concentrations and fluxes [65]. A HAPS could similarly host onboard payloads, e.g., a gas chromatography or laser spectrometer unit drawing ambient air, to continuously monitor trace gases over a fixed location for months. Moreover, a HAPS can act as a deployment platform for flying sub-sensors: it might release dropsondes or coordinate with drones in a "mothership" role to sample areas of interest below. Aside from onboard sampling, "flying sensor" here also means using an aircraft/HAPS as a mobile remote sensing platform akin to how NASA research aircraft operate. The HAPS can carry downward-looking imaging spectrometers, thermal cameras, or lidars to scan the surface and lower atmosphere at will. The benefit of using aircraft/HAPS for sensing is the ability to obtain very high spatial resolution and flexibility in targeting, at the cost of limited coverage area. For instance, airborne remote sensing can achieve centimeter–level spatial resolution for imaging (far better than satellites), which is ideal for local environmental surveys (e.g., mapping a methane leak in a facility or monitoring algal blooms in a lake). HAPS in particular offer a unique middle ground: they can hover over one region, providing continuous coverage (unlike a satellite's brief pass) and cover a broader area for longer durations than a piloted aircraft. The trade-off, however, is that a single HAPS has limited spatial coverage relative to satellite networks; essentially, it is constrained to the region within line-of-sight. Thus, HAPS airborne sensing excels in regional observations at high resolution, whereas satellites provide global coverage at coarser resolution [66].

### E. Differential Absorption Lidar (DIAL)

DIAL is an active laser remote sensing method that measures atmospheric gas concentrations by exploiting wavelength-specific absorption. In a DIAL system, the transmitter emits laser pulses at two (or more) wavelengths: one tuned to a strong absorption line of the target gas ("on-line") and one nearby where the gas has little or no absorption ("off-line") [67]. As these laser pulses propagate through the atmosphere and scatter back to the receiver (via molecules and aerosols), they experience differential attenuation due to the presence of the target gas. By comparing the backscattered signal strength at the on-line vs. off-line wavelength as a function of range, the DIAL system can infer the concentration of the gas along the path. In essence, the on-line beam "sees" the gas absorption plus background losses, while the off-line beam sees only the background losses; subtracting the two gives the gas absorption component [67]. DIAL (when performed using a pulsed laser) is a form of lidar, so it provides range-resolved information (e.g., a vertical profile of gas concentration) by timing the return signals. It is a well-established technique for measuring trace gases like ozone, water vapor, $CO_2$, $SO_2$, etc., with high sensitivity and specificity. Notably, DIAL is self-calibrating: using a ratio of two wavelengths cancels out many systematic factors like beam divergence and baseline mirror reflectivity. A classic example is ozone DIAL: a UV laser at ~289 nm (ozone-absorbing) and ~299 nm (non-absorbing) can profile ozone from the surface to upper atmosphere. DIAL can be deployed on HAPS. offering day/night operation (active illumination) and high precision. Fig. 4 shows a schematic of a HAPS-based DIAL with an example of on/off-line pair selection. Retro-reflector-in-space (RIS) sensing is a differential-absorption-lidar variant that replaces the usual aerosol/ground backscatter with a specular return from a passive corner-cube mirror carried by a LEO satellite, forming an Earth-satellite-Earth (or potentially Earth-HAPS-Earth) long-path cell for gas absorption spectroscopy. Flight demonstrations on Japan's ADEOS platform in the 1990s confirmed that a 0.5 m hollow cube could survive orbit and deliver usable methane- and ozone-absorption signals to a 1.5 m ground telescope [68,69], and follow-up dual-TEA-$CO_2$ experiments retrieved trace-gas columns over 1,000 km slant ranges [70,71]. However, classical monostatic DIAL remains the more practical choice for today's HAPS payloads: its diffuse molecular/ground backscatter tolerates milliradian pointing errors, needs only a single telescope, and already supports range-resolved profiling as shown by NASA's airborne HALO and numerous UAV campaigns, whereas RIS demands sub-arc-second beam steering, continuous satellite tracking, and offers only integrated-column data. The passive corner-cube mirror eliminates space-borne lasers but shifts a stringent pointing burden onto the platform and adds orbital-mechanics constraints (limited overpass windows) that complicate continuous monitoring.

Recent differential-absorption-lidar (DIAL) work spans every flight tier. Early balloon studies by Heaps et al. proved that stratospheric DIAL from ~30 km is feasible, setting the template for modern HAPS concepts [72,73]. At the aircraft level, NASA's High-Altitude Lidar Observatory (HALO) flew a 1.645 µm $CH_4$ DIAL/IPDA on a Gulfstream-III in 2023, retrieving city-scale methane columns during the STAQS campaign [74]. A pulsed 1.6 µm system from NASA LaRC has been rigorously tested on ground, crewed aircraft, and small UAVs, demonstrating <1 % column-error methane sensing and thus extending DIAL to drone altitudes [75]. Looking upward, the Franco-German MERLIN minisatellite, slated for a 2029 launch, will carry an IPDA lidar to deliver global $CH_4$ maps at 50 km horizontal resolution from 500 km orbit, effectively taking DIAL into space [76]. Table III summarizes representative remote sensing experiments (with focus on DIAL) across different platforms.

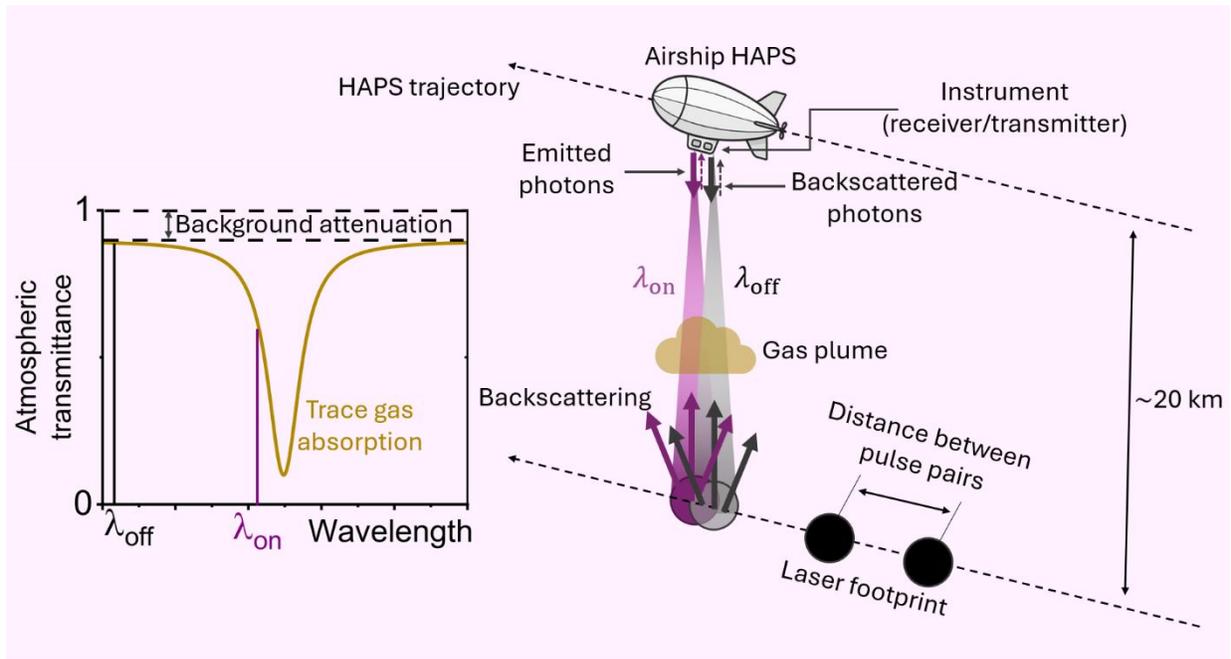

**Fig. 4.** Schematic representation of a dual-frequency DIAL system.

TABLE III
LASER-BASED REMOTE SENSING STUDIES IN LITERATURE.

| Platform | Laser / radiation source | λ (µm) | Species | Spatial resolution | Detection sensitivity | Technique | Altitude (km) | Ref. |
|---|---|---|---|---|---|---|---|---|
| Aircraft | Passive SWIR solar-backscatter imaging spectrometer | 1.56 – 1.69 | $CO_2$ | ≈ 50 m × 63 m | 0.30 % at fixed point | Passive DOAS / imaging spec. | 3 | [75] |
| Aircraft | Locked diode-seeded pulsed IPDA lidar | 1.572 | $CO_2$ | - | 0.7 ppm (1-sec averaging) | Pulsed IPDA lidar | 3 - 12 | [77] |
| Aircraft | Intensity-modulated CW diode lidar | 1.57 | $CO_2$ | - | 1.3 – 2.2% (0.1-sec averaging) | Intensity-modulated continuous-wave IPDA lidar | 8 - 12 | [78] |
| Aircraft | Ho:Tm:YLF double-pulsed laser | 2.05 | $CO_2$ | 0.75 m | 1.02% | Pulsed IPDA lidar | 6 | [79] |
| LEO satellite (pseudo measurements/simulations) | Diode-seeded IPDA | 1.572 | $CO_2$ | 50 km | 0.59 ppm systematic (16-day) | Space-borne IPDA (sim.); Aerosol and Carbon Detection Lidar | 705 | [80] |
| LEO satellite (MERLIN), launch planned in 2029 | Nd:YAG-pumped OPO IPDA | 1.645 | $CH_4$ | 50 km | Planned: ≤ 3.7 ppb | Space-borne IPDA lidar | 500 | [81] |
| LEO satellite | Dual TEA-$CO_2$ lasers | ~10.0 | $O_3$ | - | 2% | Differential absorption with RIS | 797 | [82] |
| LEO satellite concept (HSR-lidar) | Nd:YAG | 0.532 | Aerosol / cloud | 5 km | ≤ 40 % (sim.) | High-spectral-resolution lidar (Mie and Rayleigh scattering) | 705 | [83] |
| Aircraft | Earth thermal radiation | 3.7 -15.5 | CO | 50 m | ppb levels | Passive infrared interferometry | 20 | [84] |

| Platform | Laser / radiation source | λ (μm) | Species | Spatial resolution | Detection sensitivity | Technique | Altitude (km) | Ref. |
|---|---|---|---|---|---|---|---|---|
| Low-angle ground-based probing of upper atmosphere | OPO* pumped at 1064 nm | 1.6022 ($CO_2$); 1.6455 ($CH_4$) | $CO_2$ and $CH_4$ (mixing ratios) | ≈ 3 km (at 22 km) | 1% | DIAL | 22 | [85] |
| Horizontal path in the atmosphere | MIR OPO | 3.3 – 3.5 | $CH_4$ | 100 | 13% | DIAL | 0.8 | [86] |
| Ground-based | Nanosecond Ti:sapphire laser | 0.77 | $O_2$ | - | < 1 ppm·km | DIAL | 0 | [87] |

### F. Compatibility with HAPS and ISAC

Table IV compares various remote sensing techniques and assesses their compatibility with HAPS.

TABLE IV

COMPARISON OF REMOTE SENSING TECHNIQUES BASED ON THEIR DEPLOYMENT PLATFORMS, MEASUREMENT PRINCIPLES, STRENGTHS, AND LIMITATIONS, WITH A SPECIFIC FOCUS ON THEIR COMPATIBILITY WITH HAPS.

| Technique | Typical Platforms | Measurement Principle | Key Strengths | Key Limitations | Compatibility with HAPS |
|---|---|---|---|---|---|
| **Solar occultation** | HAPS balloons, satellites | Record direct sunlight while the Sun grazes the atmospheric limb at dawn/dusk; retrieve vertical profiles from differential absorption | Very high vertical resolution (~1 km) and SNR | Only two opportunities per orbit/day; limited spatial coverage; requires precise Sun alignment | ✓ |
| **Scattered-sunlight (nadir / limb) spectroscopy** | HAPS, drones, aircraft, satellites | Analyse sunlight that has been reflected (nadir) or scattered sideways (limb) to infer gas columns | Broad daytime coverage; enables large-area mapping | Coarser vertical info; cloud / surface-albedo effects; needs radiative-transfer modelling | ✓✓ |
| **Heterodyne (coherent) detection** | Doppler-wind lidars on aircraft; high-resolution spectrometers | Mix received optical signal with a local-oscillator (LO) laser on a photodetector; beat-note (IF) frequency encodes Doppler shift or spectral detail | Ultra-high spectral / velocity precision (< 1 m s$^{-1}$); phase information preserved | Requires stable LO laser, turbulence mitigation, and more complex hardware | ✓✓ |
| **In-situ / onboard sensors** | HAPS, UAVs, research aircraft | Directly sample ambient air with gas analysers, particle counters, dropsondes, etc. | Ground-truth accuracy; very high spatial/temporal resolution | Limited to platform altitude / coverage; payload mass & power constraints | ✓✓✓ |
| **DIAL** | Drones, aircraft, HAPS, satellites | Transmit paired "on-line" & "off-line" laser pulses; differential backscatter reveals range-resolved gas absorption | Self-calibrating; works day/night; high-precision range profiles | Needs powerful tunable lasers; eye-safety constraints; aerosol interference | ✓✓✓ |

DIAL uniquely pairs self-calibrating, range-resolved gas retrievals with the narrow, eye-safe laser lines already common in free-space-optical (FSO) links. After four decades of development, the technique has reached high technology-readiness as demonstrated by various applications. These milestones confirm a mature supply chain of tunable diode-pumped lasers, high-throughput telescopes and photon-counting detectors that can be down-scaled for high-altitude-platform stations (HAPS). Critically, DIAL's transmitter–receiver optics, beam-steering gimbal and fine-tracking sensors mirror the hardware already needed for gigabit FSO backhaul; recent laboratory prototypes even show single-aperture schemes that time-share lidar pulses and communication traffic without performance loss [88]. By capitalizing on this hardware congruence, a HAPS payload can deliver simultaneous broadband connectivity and quantitative greenhouse-gas sensing with minimal mass and power overhead, i.e., ISAC. For these reasons (proven accuracy, mission heritage, and intrinsic compatibility with FSO terminal design) this paper adopts DIAL as the primary recommended sensing modality. The next section will delve into the fundamentals of how DIAL quantitatively retrieves gas concentration from the two-wavelength measurements.

## VI. DIAL Fundamentals

DIAL operates on the principle of the Beer–Lambert law applied in a range-resolved lidar context. The power $P$ received from a distance $R$ for a single-wavelength lidar is given (in simplified form) by:

$$P(\lambda, R) = \frac{P_0(\lambda) C}{R^2} \exp\left[-\int_0^R \alpha(\lambda, r) \, dr\right], \quad (1)$$

where $P_0$ is the transmitted power, $C$ encompasses the telescope and detector gain, and $\alpha$ is the atmospheric attenuation coefficient at wavelength $\lambda$ (due to both the target gas absorption and other extinction like aerosols). In a DIAL measurement, we have two wavelengths $\lambda_{on}$ and $\lambda_{off}$ chosen such that the target gas has an absorption cross-section ($\sigma_{on}$) significantly higher at the on-line than at the off-line ($\sigma_{off}$) (where it is negligible). Over a small range segment, the difference in optical depth between the two wavelengths is dominated by the target gas. By taking the ratio (or difference in the log domain) of the on vs. off returns, the common path effects (like aerosol scattering and geometrical $R^2$) cancel out to first order.

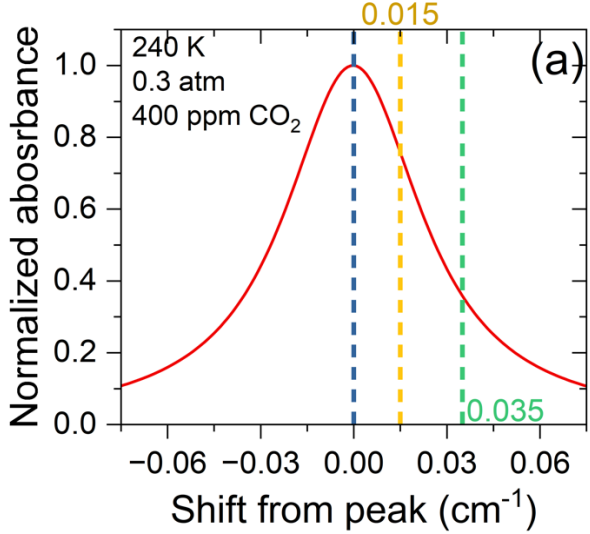
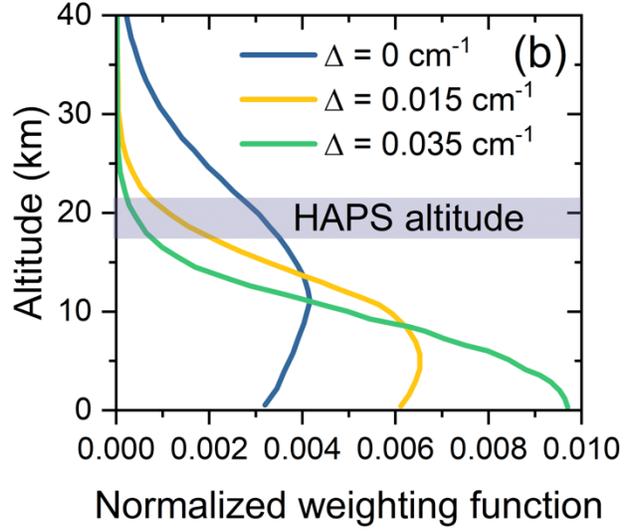

**Fig. 5.** (a) Normalized absorbance profile for $CO_2$ showing proposed on-line wavelength positions at several spectral shifts from the line-center (6361.248 cm⁻¹). (b) Corresponding normalized weighting functions, reproduced from [80].

Quantitatively, the gas number density $N(r)$ at range $r$ can be derived from the range-derivative of the log-power ratio between off-line and on-line signals:

$$N(r) \approx \frac{1}{2(\sigma_{on} - \sigma_{off})} \frac{d}{dr} \ln\left[\frac{P_{off}(r)}{P_{on}(r)}\right]. \quad (2)$$

In essence, the slope of the logarithmic range-corrected power ratio is proportional to the gas concentration. A more practical form often used is to compute the average $N$ in a layer between two range points $r_1$ and $r_2$ from the difference in signals at those ranges. For example, one can rearrange the above equation to:

$$N_{avg}(r_1, r_2) = \frac{1}{2(\sigma_{on} - \sigma_{off})} \ln\left[\frac{P_{off}(r_2) P_{on}(r_1)}{P_{off}(r_1) P_{on}(r_2)}\right]. \quad (3)$$

This approach uses the so-called DIAL equation, which has been well documented in the literature. The factor of ½ appears because the two-way (out-and-back) optical depth is considered. A commonly used expression which introduces a weighting function ($WT$) to calculate absorbance, $A$, accounting for temperature ($T$) variations with altitude is given as:

$$A = \ln\left(\frac{P_{off}}{P_{on}}\right) = 2 \int_{r_1}^{r_2} WT(r, T) \cdot N(r) dr. \quad (4)$$

The weighting function primarily depends on the differential absorption cross-section of the target molecule and the air's molecular number density. When atmospheric conditions are fixed, its profile varies solely with the selected laser wavelengths. For instance, the weighting function for measuring $CO_2$ columns can be expressed as:

$$WT(r, T) = \frac{\sigma_{CO_2,on}(r,T) - \sigma_{CO_2,off}(r,T)}{(m_{H_2O} N_{H_2O}(r) + m_{dry}) g}, \quad (5)$$

where $m_{H_2O}$ and $m_{dry}$ are, respectively, the average mass of water and air molecules and $g$ is the gravitational acceleration. The column-weighted dry air mixing ratio of $CO_2$ can then be determined as

$$X_{CO_2} = \frac{A}{2 \int_{r_1}^{r_2} WT(r,T) dr}. \quad (6)$$

While the absorption line-center (peak) provides maximum detection sensitivity, it is recommended to slightly shift the online wavelength from the absorption peak. This shift avoids potential absorption saturation (too much laser attenuation) as well as makes DIAL less sensitive to laser wavelength drift and variations to absorption line-shape due to temperature variations. Wang et al. [89] characterized the $CO_2$ line centered at 6361.248 cm⁻¹ for DIAL applications. Fig. 5a shows simulation (HITRAN2020 [90]) of this $CO_2$ absorption line at average conditions with two peak shifts proposed for the on-line measurements. Fig. 5b shows the calculated weighting function corresponding to the line peak ($\Delta = 0$) and the two proposed shifts, obtained from Wang et al. [80]. Fig. 5b illustrates how the normalized weighting-function profiles change with both altitude and the chosen on-line wavelength. Higher values of the weighting function indicate altitudes where the DIAL is most responsive to atmospheric variations. Because identifying $CO_2$ sources and sinks hinges on strong tropospheric signals, a space-borne DIAL must therefore maximize its sensitivity within that lower-atmosphere layer. Evidently, shifting the on-line wavelength away from line-center flattens the weighting-function peak, reducing upper-tropospheric dominance and improving sensitivity to lower altitudes.

It is important to note some assumptions in the ideal DIAL equation: it assumes that other attenuators (aerosols, Rayleigh scattering) affect both wavelengths equally, so they cancel out in the ratio. In reality, $\lambda_{on}$ and $\lambda_{off}$ are chosen to be as close as possible (while still having different gas absorption) in order to minimize differences in aerosol scattering. Residual differences (for instance, if one wavelength is slightly more aerosol-sensitive) can introduce systematic error. DIAL systems often incorporate corrections or calibrations, e.g., using an iterative correction with an assumed aerosol backscatter profile. Fortunately, in many cases of gas sensing, the absorption feature can be found in a spectral "window" where aerosol extinction is smoothly varying. Experiments have shown that if wavelengths are well-chosen, the errors from aerosol differences are low (on the order of a few percent

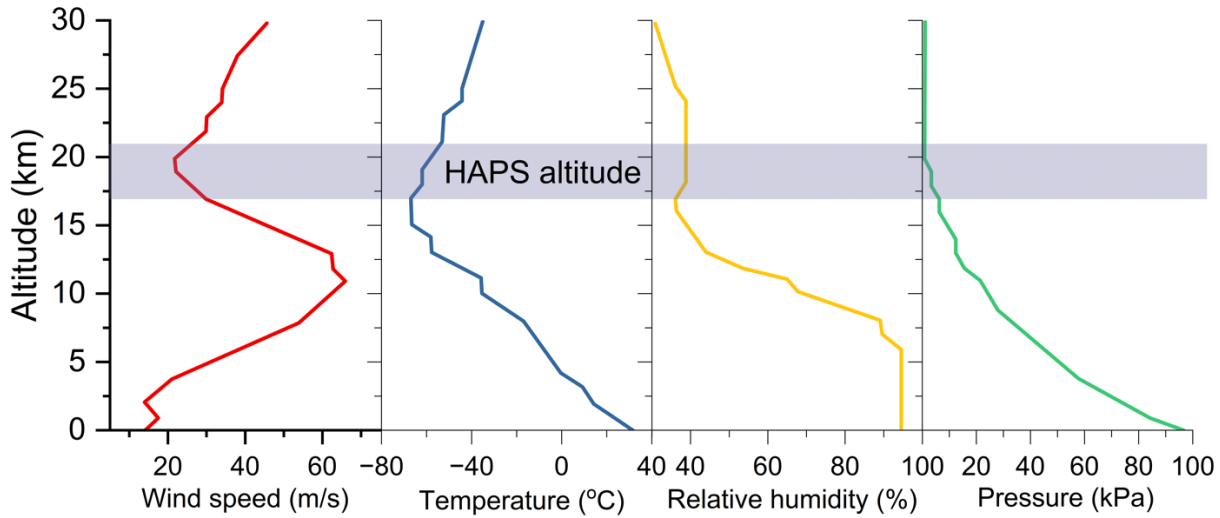

**Fig. 6**. Climatic conditions across various altitudes. The data used to generate the curves were adapted from Ref. [91].

in the lower troposphere, and negligible in the clear upper troposphere).

It is important to vertically resolve the DIAL weighting function for each target gas as shown for $CO_2$ in Fig. 6. A radiative-transfer model numerically solves the radiative transfer equation along the full slant path, accounting for altitude-dependent profiles of temperature, pressure, and humidity, which are shown in Fig. 6 (99th percentiles of conditions, data adapted from Ref. [91]). It incorporates absorption based on the Beer–Lambert law, as well as emission and, when relevant, scattering processes. This model is used to forward-simulate both on-line and off-line signals, particularly for remote sensing applications such as DIAL.

Because winds in the lower stratosphere dip to a broad minimum as shown in Fig. 6, this layer is described as the HAPS sweet spot, offering a relatively calm environment that minimizes structural loads, station-keeping power, and beam-jitter noise. As the laser descends it enters windier, more turbulent tropospheric layers; here fast-steering mirrors and adaptive tip–tilt control are needed to stabilize the on/off beam overlap and to keep differential-power ratios within calibration tolerance. Temperature and pressure gradients shape the molecular number density and pressure broadening of the absorption line: dense, warm air near the surface both increases absorption and widens the Lorentz wings, whereas the cold, rarefied stratosphere leaves a narrow Doppler core whose magnitude depends sensitively on temperature. Relative humidity drops significantly above 8 km, so the upper portion of the path contributes negligible $H_2O$ interference, yet the moist boundary layer can still imprint extra attenuation; wavelength selection therefore should avoid major $H_2O$ lines.

It should be noted that there are two modes of DIAL operation: range-resolved DIAL (RR-DIAL) and integrated-path DIAL (IPDA). In RR-DIAL, one relies on the atmospheric backscatter at the wavelength itself (usually from aerosols or molecules) to get a return from every slice of the atmosphere. In IPDA, one uses a distant hard target (like the ground or a cloud) as the reflector; the measurement yields the total column amount of gas between the instrument and that target [92]. IPDA is often used for greenhouse gas sensing from high altitudes or space, where the ground return is strong. The differential transmission between on/off over the entire path gives the column-average mixing ratio [92]. A HAPS-based IPDA lidar could, for instance, ping the ground with on/off pulses to measure the column $CO_2$ or $CH_4$ below the platform (covering the lower stratosphere and troposphere). In contrast, a RR-DIAL on HAPS might focus on profiling a specific altitude range (e.g., water vapor in the boundary layer) by analyzing backscatter within that region. Both modes share the same core principles of differential absorption.

An important aspect for RR-DIAL is the range resolution. Because it relies on differentiating the signal, the effective resolution is often coarser than the raw LIDAR range bin spacing. Some form of data smoothing or grouping of spatial bins is needed to get a usable signal-to-noise ratio in the derivative. Thus, a DIAL might achieve vertical resolution on the order of tens of meters to a few hundred meters, depending on signal strength and optical depth of the target gas [93]. Despite this, DIAL's ability to directly provide range-resolved concentration profiles is unparalleled for certain gases. For example, airborne DIAL measurements of water vapor have provided humidity profiles with high precision and accuracy, addressing a longstanding need in meteorology [94]. Likewise, ozone DIALs have mapped stratospheric ozone layers in 3D during research campaigns.

## VII. ROLE OF HAPS IN DIAL-BASED ENVIRONMENTAL MONITORING

HAPS offer a novel advantage point and operational mode for environmental monitoring, especially when equipped with DIAL and other remote sensing instruments. They effectively bridge the gap between satellites and ground-based observations, enabling unique applications and providing several advantages.

### A. Persistent Regional Monitoring

Perhaps the greatest strength of HAPS is their ability to loiter over one area for extended periods. This is invaluable for tracking environmental phenomena that evolve on timescales of hours to weeks. For example, a HAPS with a DIAL could continuously monitor the air quality over a city, measuring diurnal buildup and dispersal of pollutants ($O_3$, $NO_2$, particulate extinction) in a way satellites (with once-to-few-per-day overpasses) cannot. The continuous, real-time observation of critical indicators, like pollution plumes, forest fire smoke, volcanic ash, or greenhouse gas leaks, allows for early warning and responsive mitigation. During the 2020s, ecological crises (wildfires, industrial accidents releasing gases, etc.) have underscored the need for rapid situational awareness. HAPS carrying advanced sensors can be rapidly deployed to such hotspots and provide high-fidelity data in real time [95]. For instance, in the event of a large methane or toxic gas leaks, e.g., $H_2S$, from an infrastructure site, a HAPS-based DIAL could map the concentration plume and quantify emission rates continuously, enabling operators and regulators to take informed action. This persistent monitoring capability is something neither satellites (brief snapshots) nor ground stations (point measurements) alone can achieve. It effectively creates a "virtual sensor tower" 20 km high that oversees the region.

### B. High Resolution and Low Altitude Advantage

Operating at ~20 km altitude, HAPS observe the Earth from much closer range than satellites (which at best are < 400 km for very low Earth orbit satellites (vLEO)). This yields finer spatial resolution and stronger signal returns. Lidar and imagery from HAPS can attain very high resolutions (on the order of meters to tens of meters) over a wide swath, which is extremely useful for local environmental applications. For example, a HAPS DIAL could scan a methane plume with a horizontal resolution of a few tens of meters, distinguishing individual source areas in an oil/gas field; something that space-based sensors with ~0.5–1 km resolution would average out. Moreover, the shorter range means less atmospheric attenuation and higher signal-to-noise, allowing detection of trace gases in lower concentrations. Studies have noted that HAPS can complement satellites by providing high-resolution data to validate and refine the coarser satellite maps [48,96]. HAPS can also venture below cloud decks or target specific altitudes of interest by adjusting flight level (somewhat like adjustable "viewing geometry"), which satellites cannot do. In essence, HAPS add a mesoscale observing layer in the Earth observation hierarchy, covering areas of order 10–100 km with great detail and flexibility [96].

### C. Flexibility and Rapid Deployment

Unlike satellites that require long lead times and fixed orbits, HAPS can be launched or redirected on-demand. This makes them excellent for emergency environmental monitoring, e.g., tracking the spread of hazardous emissions from an industrial accident or providing surveillance of an oil spill's atmospheric impact. HAPS can be on station within hours (if pre-positioned at an airport) and relocated as needed. Their deployment is also reversible (they can be landed for maintenance or payload swap). This flexibility allows using the right sensors at the right time: a HAPS might carry a generic payload most of the time, but if a volcano erupts, one could equip it with an $SO_2$ DIAL and move it nearby to measure the eruption's gas output continuously. Indeed, international agencies have recognized the importance of HAPS for disaster support. They can restore communications and provide remote sensing in disaster zones when ground infrastructure is compromised [95]. For example, after wildfires, a HAPS could monitor air quality and smoke dispersion to guide public health advisories. After a flood or tsunami, it might map the extent of standing water and potential contamination. This on-demand re-tasking is a significant advantage over orbiters, which cannot adjust their trajectory for specific events.

### D. Synergy with Satellites and Ground Networks

HAPS are best thought of not as replacing satellites or ground sensors, but as augmenting them in a multi-tier "observation pyramid." They fill the gap of intermediate scale observations. Environmental monitoring often benefits from data at multiple scales: satellites give the big picture, ground stations give point verification, and HAPS can intermediate with regional context [48,96]. For instance, in climate research, to understand $CO_2$ sources and sinks, one strategy is to use a satellite (like OCO-2 or TanSat) to identify regional anomalies, then deploy a HAPS DIAL to hover over that region (like a rainforest or city) for weeks to directly measure fluxes and vertical profiles, then feed that information back to refine global models. HAPS data can also help calibrate and validate satellite sensors: having a long-duration DIAL in the same air column that a satellite passes over provides quasi-coincident measurements to check the satellite's accuracy. Researchers have pointed out the "unmet observational requirement" for platforms that can loiter and follow air masses, working in concert with spaceborne instruments [48]. HAPS uniquely satisfy this by being able to stay with an airmass as it moves (they can reposition slowly with wind/propulsion), for example, a HAPS could follow a pollution plume as it drifts hundreds of kilometers, something neither fixed stations nor orbiters could continuously do. This opens new possibilities for studying atmospheric transport and chemistry in a Lagrangian frame (moving with the air).

*E. Wide Applications and Use Cases*

HAPS-based DIAL offers a broad spectrum of potential applications. Fig. 7 summarizes some prominent examples which are discussed further in this section.

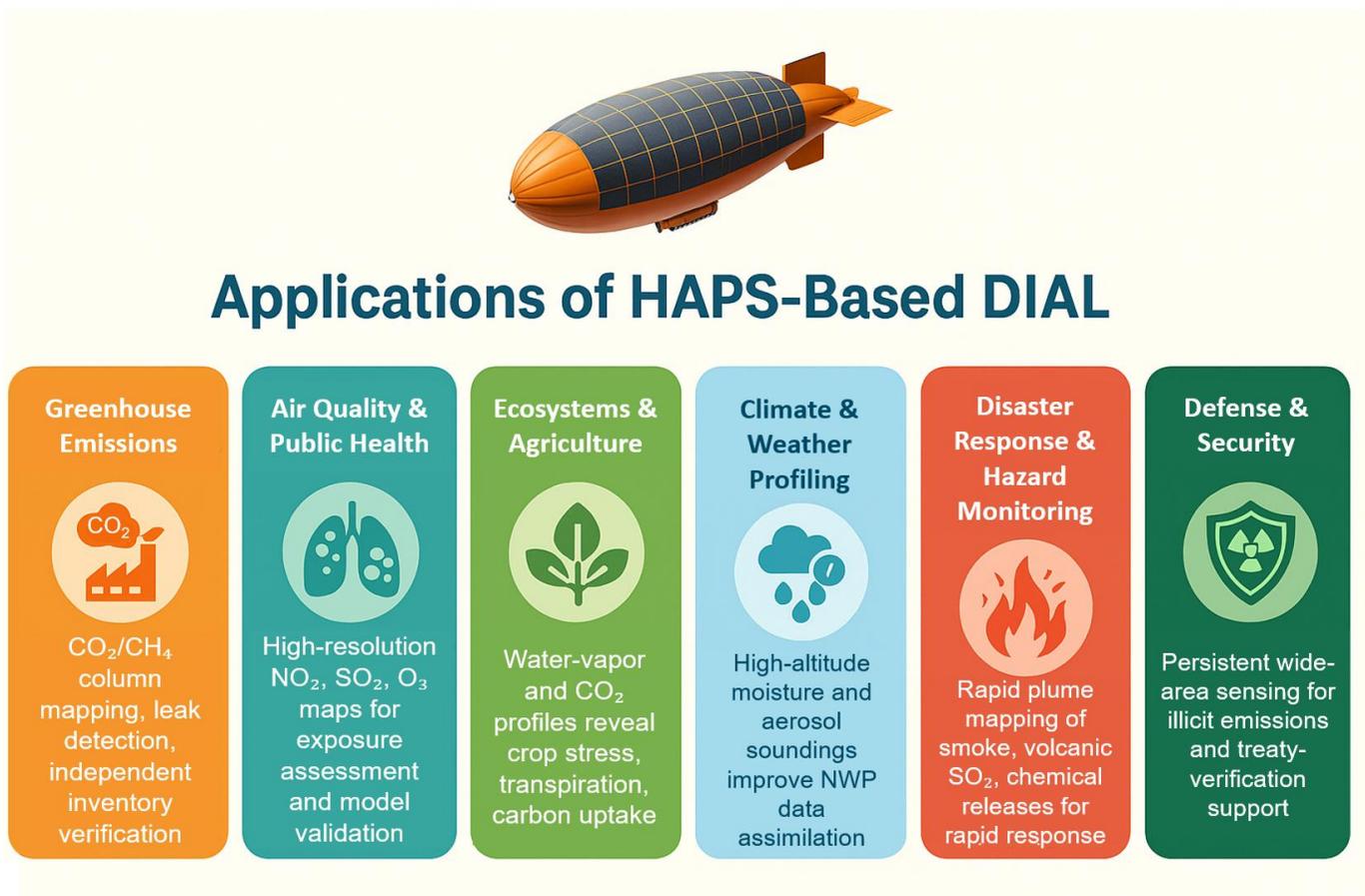

**Fig. 7.** Potential applications of HAPS-based DIAL for environmental monitoring; NWP: Numerical Weather Prediction.

*1) Greenhouse Gas Monitoring:* HAPS can play a role in climate change monitoring by quantifying greenhouse gas emissions. They can surveil major industrial or natural sources, e.g., measure $CO_2$, $CH_4$, and $N_2O$ above large cities, power plants, wetlands, or permafrost regions. Because they can hover, they are well-suited to measure fluxes: one can imagine stationing a HAPS upwind and downwind of a metropolitan area to directly measure how much $CO_2$ the city is emitting by differential concentration analysis (an analog of tall towers but covering a larger scale). Such data are crucial for verifying emission inventories and detecting super-emitters. In fact, a HAPS with $CH_4$ DIAL could systematically scan oil and gas basins for gas leaks, complementing satellite detections with finer resolution and persistence. While MERLIN (the satellite-based $CH_4$ lidar) will provide global context, HAPS can focus on regions flagged as problematic, enabling continuous localized monitoring. The HAPS Alliance and other organizations have highlighted environmental monitoring, specifically greenhouse gas and air quality tracking, as a key high-impact use case for these platforms [97].

*2) Air Quality and Public Health:* Urban air pollution (PM2.5, ozone, NOx, etc.) is typically monitored by sparse ground stations. A HAPS could provide 3D mapping of pollutants over a city: for example, a DIAL tuned to aerosol backscatter at 355 nm (for particulates) or to $NO_2$ absorption could give vertical distribution of smog. This helps identify pollution layers (e.g., nocturnal inversion layers trapping emissions) and can feed more accurate data to health alerts and traffic regulation systems. Because HAPS can stay for days, they capture the full cycle of pollution build-up and dispersal through weather changes. This persistent 3D data could improve high-resolution air quality forecasting. HAPS can also facilitate "chemical weather" observation by combining different sensors (one platform might carry a DIAL for ozone and a hyperspectral imager for $NO_2$, for instance). City authorities could deploy HAPS in the event of severe pollution episodes (e.g., wildfire smoke affecting a city) to assess the vertical extent and concentration of hazardous smoke, guiding advisories for mask-wearing or evacuation for sensitive groups.

*3) Ecosystem and Agricultural Monitoring:* With appropriate sensors (multispectral cameras, DIAL), HAPS can monitor forests, crops, and water bodies. For environmental management, a HAPS could map vegetation health via hyperspectral imaging and also measure atmospheric variables like $CO_2$ uptake in the area via DIAL transects. This coupling of biosphere observation with atmosphere measurements could reduce uncertainties in carbon budgeting for ecosystems. Future developments may involve networking HAPS with ground-based IoT sensors (e.g., soil moisture probes, flux towers), where the HAPS would simultaneously provide connectivity and overhead monitoring, thereby integrating sensing, communication, and edge computing for smart agriculture [95].

*4) Climate and Weather Research:* Water vapor is a key driver in weather and climate, and is notoriously under-sampled (weather balloons are launched typically 2x daily). A HAPS equipped with a water vapor DIAL could continuously profile humidity in a critical region (e.g., above the ocean in hurricane genesis areas, or in the Arctic atmosphere to study moisture intrusions). This data could be assimilated into weather models to improve forecasts. Also, HAPS can measure winds (with Doppler lidar) and temperature (with IR absorption or microwave sensors), functioning almost like a "moveable weather station" high in the atmosphere. NASA and other agencies have considered HAPS in Earth observing systems to fill temporal and spatial gaps; for instance, in the absence of a geostationary satellite over polar regions, a HAPS could provide continuous coverage [48]. In the coming years, we may see dedicated HAPS for meteorology, flying around to sample developing storms or to maintain constant watch over vulnerable regions (similar to how National Oceanic and Atmospheric Administration (NOAA) uses drones and aircraft now, but with far longer endurance and at higher vantage). The fact that HAPS can stay above most weather (at 20 km, above tropospheric clouds) means their remote sensors have a mostly clear view of the lower atmosphere, improving data availability compared to satellite-to-ground measurements that can be obscured by clouds.

*5) Defense and Security Applications:* Military and defense users are likewise keen to exploit multi-mission HAPS that can serve as ISR assets while maintaining robust communication links. A single high-flying platform could carry electro-optical/infrared cameras, LIDARs, or chemical sensors to surveil a wide area, and simultaneously act as a node to relay the gathered intelligence to commanders or other platforms via laser communications. The appeal of optical (laser) communication in defense is its high bandwidth and inherent low probability of interception; a narrowly directed laser beam is extremely difficult for an adversary to detect or eavesdrop on without being in its direct path [56]. This provides secure, jam-resistant backhaul for ISR data, in contrast to traditional RF datalinks that are susceptible to interception and jamming. By integrating an optical communications terminal and, say, a laser-based chemical/biological sensor on the same HAPS, the military can detect potential threats (e.g., a toxic gas release or a nuclear material signature) and instantly send encrypted data about these threats to ground stations or airborne command posts. Such capability is valuable for CBRN (chemical, biological, radiological, nuclear) defense and remote sensing of battlefields or borders.

## VIII. INTEGRATED SENSING AND COMMUNICATION (ISAC) ON HAPS

Reliable, high-capacity communications are the linchpin that turns HAPS from into a strategic asset. Operating 18–25 km above weather and air traffic yet far closer than satellites, a HAPS can project gigabit-class links over a 100-km radius with fibre-like latency, stitching remote communities into national backbones, backhauling 5G/6G small cells, and giving emergency responders instant network coverage after disasters. Persistent line-of-sight to ground stations, airborne assets, and even LEO satellites enable a single platform to function as a bidirectional relay: aggregating sensor data, coordinating unmanned aerial swarms, or offloading traffic from congested terrestrial networks. Because SWaP (size, weight and power) budgets are tight, every added sensing or navigation payload depends on a robust communications subsystem for command-and-control, time synchronization, and high-rate data exfiltration. Consequently, a wide array of both experimental and theoretical initiatives has been undertaken to enable robust communications on HAPS platforms. Table V summarizes some of these recent efforts.

TABLE V
A LITERATURE SURVEY OF HAPS COMMUNICATION SYSTEMS.

| Platform | Spectral Band | System Description | Max Data Rate / Bandwidth | Operational Altitude | Coverag/ Link Range | Ref. |
|---|---|---|---|---|---|---|
| Airship | 28/31 GHz & 47/48 GHz mm-wave access links + free-space optical backhaul / inter-platform links | EU FP6 *CAPANINA* project: develops a low-cost broadband HAP network that combines smart-antenna mm-wave user links with optical backhaul to serve hard-to-reach communities and high-speed trains | 120 Mbit/s | 17 - 22 km | ~60 km | [98] |
| Balloon | 1550 nm free-space optical downlink (beacons at 810 nm & 978 nm) | The stratospheric optical payload experiment (*STROPEX*): Free-space Experimental Laser Terminal (FELT) sending a broadband optical backhaul to a transportable optical ground station | 1.25 Gb/s | 22 km | ~64 km | [99] |
| Solar-powered fixed-wing aircraft | 28 GHz and 47/48 GHz mm-wave bands allocated to HAPs (ITU); | EU *HeliNet* project: a cellular broadband payload with an array of aperture antennas on the aircraft forming 121 hexagonal spot-beams (6 km diameter each); link budgets | Down-link rates (12.5 MHz channels) up | 17 - 20 km | ~60 km | [100] |

| Platform | Spectral Band | System Description | Max Data Rate / Bandwidth | Operational Altitude | Coverag/ Link Range | Ref. |
|---|---|---|---|---|---|---|
| | lower-band 2 GHz considered only for UMTS, not for the broadband payload | and interference managed via adaptive reuse and rain-fade mitigation | to 40 Mb/s per user | | | |
| HAPS (general) | Hybrid: millimetre-wave E-band (≈ 60–80 GHz) feeder links (10 GHz per link with dual polarization) + 1550 nm free-space-optical feeder link to a GEO satellite; user service links may reuse Ka-, Q- or Ku-bands from the satellite | The balloon (HAP 140) aggregates dozens of concurrent high-capacity RF feeder links from a compact ground terminal array, converts them to a single optical uplink to the GEO satellite, and performs the reverse conversion for downlink, creating a radio/optical "hybrid" backhaul that relieves feeder-band congestion while the satellite's spot-beams serve end-users | Link budgets illustrate ≈ 1 Tb/s forward feeder capacity (≥ 40 feeder terminals) and ≈ 500 Gb/s reverse capacity at 1.25 b/s/Hz spectral-efficiency | ~17–22 km | - | [101] |
| HAPS (general) | mmWave | A steerable-beam system that lets a HAP provide both access and backhaul links to terrestrial 5G/LTE nodes, selecting links dynamically from real-time location data. | ~1 Gb/s (assumed) | ~20 km | - | [102] |
| HAPS (general) | Optical 1.55 µm | Theoretical; analytical model of HAPS-to-ground, ground-to-HAPS and HAPS-to-HAPS FSO links; studies turbulence, pointing error & adaptive-optics mitigation to predict outage probability | - | 5 - 28 km (simulated) | - | [36] |
| HAPS (Hybrid Network with ground and satellites) | Optical 1.55 µm | Analytical outage-probability model for ground↔HAPS, HAPS↔HAPS and HAPS↔GEO FSO links that includes turbulence, pointing error, weather attenuation and zenith-angle geometry | >1 Gb/s (optical) | 5 - 30 km (simulated) | Horizontal inter-HAPS link distance studied up to ~50 km | [103] |
| HAPS (generic HAP used with GEO satellite & terrestrial nodes) | Ka-band: 17.3–17.7 GHz (shared) + 19.7–20.2 GHz (exclusive); sub-band width = 36 MHz | Cognitive SatCom/HAP network: GEO broadcast-satellite feeder uplink, high-density fixed-satellite downlink and multiple HAPs share the same spectrum; iterative beamforming + sub-band allocation algorithm maximises sum-rate and mitigates inter-system interference | Simulations show network sum-rate up to ≈ 22 Gb/s for four hot-spot regions at 25 dBm per-antenna HAP power; each sub-band is 36 MHz (25 sub-bands total) | 20 km | HAP spot-beam diameter ≈ 20 km (≈ 10 km radius hot-spot) | [104] |
| Solar-powered stratospheric UAV (PHASA-35–class) equipped with phased-array antennas for multi-spot- | 27.5 GHz mm-wave with 200 MHz channel (authors also note 2.1 GHz IMT band as an option) | Downlink NOMA framework with joint user-grouping, beam optimization and power allocation; designed for remote areas with radius RH = 60–400 km | Simulations show peak average sum-rate ≈ 3.3 Gb/s over 200 MHz (bandwidth) at midday | 21 km (within the 18–24 km stratospheric window) | Reference scenario uses 60 km ground-radius footprint; framework scalable up to 400 km radius | [33] |

| Platform | Spectral Band | System Description | Max Data Rate / Bandwidth | Operational Altitude | Coverag/ Link Range | Ref. |
|---|---|---|---|---|---|---|
| beam service | | | | | | |
| Fixed-wing stratospheric aircraft | 5 G NR Band n1 @ 2.1 GHz (10 MHz channel) for the access link + 5.7–5.9 GHz (≤ 30 MHz) backhaul link | Regenerative architecture: full 5 G gNB and backhaul radio carried on the aircraft; trial run by Stratospheric Platforms Ltd. & Deutsche Telekom to deliver 5 G service from the stratosphere | Peak user down-link ≈ 90 Mbit/s over the 10 MHz channel | 14 km during the demonstration flights (design target 17–20 km for future ops) | ~450 km² footprint while loitering on a 12 km-radius circle | [10] |

*QinetiQ*, highlights that on military platforms with SWaP constraints, an optical system that can switch between high-speed communication and sensing modes (inspired by the versatility of software-defined radios) provides "unprecedented multi-functionality" [56,105]. They demonstrate that a HAPS-mounted laser system might one moment transmit gigabits of data, and the next moment perform ISR tasks like 3D mapping of terrain or vibrometry-based target identification using the same hardware [56]. This dual role extends to laser-based gas sensing: a HAPS can simultaneously map atmospheric conditions, such as methane or other trace signatures indicative of subsurface activity, and relay the information through a laser link, minimizing latency for actionable intelligence.

Progress and prototypes in integration are still in early stages, but promising. The idea of dual-use optical systems is gaining traction. On the academic front, a 2017 demonstration by Scotti *et al.* implemented a LIDAR architecture that could double as an FSO communication link, including a version on a photonic integrated chip [88]. This shows that in hardware terms, one can design transceivers flexible enough for both tasks. In that case, it was a coherent receiver that could detect faint LIDAR returns with high resolution, which also serves to decode optical communication signals, a clear parallel in required sensitivity and signal processing. We can expect more such dual-use designs: e.g., communication lasers that are tunable and could be used to sweep across absorption lines for spectroscopy, or large-area single-photon detectors that can both count LIDAR photon returns and serve as high-speed receivers for data (single-photon APD arrays are used in both fields).

Platform-level integration might involve combining optical communication and sensing with other HAPS payloads, like RF communications or radar. A HAPS may carry a payload suite comprising an optical communication system, an RF backup link, an optical gas sensor, and either a synthetic aperture radar (SAR) or electro-optical (EO) camera. The data from all of these could be fused onboard using AI to identify points of interest (for example, correlating a methane plume detection with an optical image of an industrial facility, then cueing the communication system to send an alert with that specific subset of data). This kind of smart, integrated payload is an active area of research. The benefit is reducing the delay between detecting something and communicating it; the HAPS could autonomously notice an anomaly and immediately initiate a high-priority data transmission about it. This is essentially bringing the concept of an "intelligent gateway" to the stratosphere, the HAPS as both sensor and router, deciding what information needs to be sent down and doing so efficiently. The integration of communications and sensing is a foundational step toward that vision.

Having discussed the motivations and technical facets, we now focus on important considerations regarding the integration of the two functionalities, high-speed optical communication and laser-based gas (or generally chemical) sensing, on the same HAPS platform. This integration can occur at multiple levels: shared hardware, shared wavelengths, combined data handling, and operational coordination.

### A. Spectral and Hardware Compatibility

Communication lasers typically operate at specific near-infrared bands optimized for minimal atmospheric loss and availability of components (e.g., 1550 nm, a common telecom wavelength, or 1064 nm in some aerospace cases). In contrast, laser-based gas sensing requires tuning to the absorption lines of target gases, for instance, ~1650 nm for methane ($CH_4$) or ~1572 nm for $CO_2$, or 935 nm for water vapor. These may or may not coincide with the ideal communication wavelength. One straightforward approach is to use separate lasers and perhaps separate apertures for the two functions. However, that duplicates a lot of hardware. A more elegant approach is wavelength-division multiplexing: using different wavelengths on a common optical train. Modern optical coatings and dichroic beam splitters could allow a telescope to be simultaneously used by a 1550 nm communication laser and a 1650 nm sensing laser, for example, sending both beams coaxially. A wavelength separator on return would direct the 1650 nm reflections to a gas sensor detector while the 1550 nm modulated beam goes to a communication receiver. This concept was validated in part by researchers who built a dual-use lidar/FSO system; their architecture, based on coherent detection, could be "easily exploited for optical communications" in addition to its sensing function [88]. Moreover, a classic DIAL payload already carries an "off-line" laser channel that sits outside the target-gas absorption feature; that wavelength experiences essentially no molecular attenuation, so it can be repurposed as a high-throughput data link without any risk of signal degradation from absorbing species and without adding an additional laser.

### B. Shared vs. Separate Apertures

If spectral multiplexing is implemented, a single aperture (telescope) could support both transmit/receive functions for communication and for DIAL sensing. This approach reduces mass and eliminates the need for two large optical heads on the HAPS. Similar designs are already used in many satellite laser communication terminals, where a single telescope handles bidirectional transmission through beam splitters; the same principle could be applied here.

However, a key challenge arises when the communication link must remain fixed on a ground station while the sensing system scans across a target area. One possible compromise is time-sharing the aperture's pointing, rapidly steering it between the ground link and the sensing targets. The feasibility of this approach depends on the system's inertia and the performance of its stabilization mechanisms.

An alternative is to mount two co-aligned telescopes on the same stabilized platform: one dedicated to the communication link and the other to scanning for sensing. This configuration increases mass but decouples the pointing requirements. Ultimately, the choice depends on mission priorities; continuous communication may necessitate dual apertures, whereas if data can be buffered or intermittent transmission is acceptable, time-multiplexing a single aperture may offer an attractive mass-saving trade-off.

### C. Data and Modulation Multiplexing

Beyond physical optics, there is also the intriguing possibility of information multiplexing. In principle, the HAPS could use its communication laser for both data transmission and sensing, embedding scientific measurements directly within the communication signal, as demonstrated in [39]. This can be achieved by modulating the laser's intensity so that it simultaneously probes gas absorption features while encoding digital information.

Another form of multiplexing occurs at the network level, where the high-speed communication link is leveraged to transmit the sensor data stream in real time. In this case, the HAPS communication system effectively serves as a telemetry downlink for the sensing payload. To implement this, the link must have sufficient bandwidth to accommodate both the primary communication payload (e.g., internet traffic or military communications) and the scientific data, which may include high-volume hyperspectral or DIAL measurements.

Given the multi-Gbps capacity of laser communication systems, such integration is typically feasible. For instance, even a high-resolution hyperspectral imager producing several hundred Mbps of data can be supported alongside other traffic. When bandwidth is limited, onboard data reduction, through compression or AI-based processing, can be employed to optimize transmission efficiency.

### D. Operational Scheduling

Integration implies temporal coexistence; the communication and sensing functions must share time and resources. A HAPS may not always be able to perform both at full capacity simultaneously, particularly when power, pointing, or bandwidth constraints overlap.

A practical solution is duty cycling or dynamic scheduling. For instance, the HAPS could prioritize communication relay during peak hours, such as daytime broadband operations when solar power is abundant, and allocate off-peak periods (e.g., nighttime) for intensive gas sensing. In event-driven scenarios, such as a detected gas leak, the system could automatically adjust its schedule to prioritize sensing, temporarily buffering communication data or switching to a backup RF link to maintain connectivity.

Such flexibility would rely on autonomous mission management software capable of making real-time decisions based on predefined priorities. For example, during a disaster response, sensing might take precedence over routine communications, whereas under normal conditions, communication throughput could remain the primary objective. In military operations, the priority balance could shift dynamically between ISR data collection and communication relay, depending on mission orders and situational needs.

### E. Interference and Cross-talk

If separate lasers are used for communication and sensing, potential interference between the two systems must be carefully considered. A gas-sensing DIAL typically relies on detecting extremely weak return signals from atmospheric backscatter or ground reflection at the laser wavelength. Concurrent operation of a high-power communication laser at a nearby wavelength could introduce optical cross-talk through stray light, backscatter, or nonlinear effects in shared optics.

Mitigation requires rigorous optical isolation. This may include high-rejection optical filters on the sensing detector, spatial separation of the two beams (e.g., small angular offsets if they do not share the exact same path), and temporal separation, ensuring that the DIAL laser does not fire during moments when the communication receiver is actively demodulating symbols, to avoid blinding or signal contamination.

These challenges are complex but solvable through coordinated system design and testing. The integrated payload should be validated under all operational modes to confirm that, for instance, firing the methane DIAL does not induce a spike in bit errors on the communication link due to scattered light entering the communications telescope. If interference is observed, synchronized operation can be implemented; for example, momentarily pausing data transmission for a few milliseconds during each DIAL pulse. With multi-Gbps communication links, such brief interruptions would result in negligible data loss, which can be fully compensated through buffering or forward error correction.

### F. Unified Data Handling

On the ground or within the wider networor, integrating data streams from a HAPS that performs both sensing and communication opens new opportunities for data fusion. For example, a HAPS could simultaneously detect a methane plume and capture a video feed; merging these data sets would produce a georeferenced video overlaid with gas concentration maps, offering users a richer situational understanding.

To enable this, the HAPS must precisely time- and geo-tag all sensor outputs. This is readily supported by the communication subsystem, which can embed timestamps within transmitted data, while onboard GPS provides accurate positional references for geolocation.

At the network level, integration continues. Ground stations receiving the combined HAPS data stream may need to route information to multiple destinations, e.g., telecommunication networks for internet traffic, and scientific or command centers for sensing data. Designing network protocols that can separate, prioritize, and deliver these data streams efficiently represents another dimension of integration, one that relies more on software architecture and data management than on hardware.

## IX. WAVELENGTH SELECTION FOR SENSING AND CO-USE WITH COMMUNICATION

Selecting an optimal wavelength is critical for remote sensing performance and for compatibility with communication system. The selection process requires balancing absorption strength (to maximize sensing sensitivity) against atmospheric transmission (to ensure efficient propagation), while also accounting for the availability of suitable lasers and detectors and avoiding spectral conflicts with other system functions. The following discussion outlines typical wavelength choices for DIAL and related sensors, and examines how these regions overlap or diverge from those used in optical communication systems.

### A. Ultraviolet (UV) and Visible

Short wavelengths (UV/visible) are often used for sensing ozone and pollutants. For example, ozone DIAL commonly uses the Huggins band in the UV around 300 nm, a classic on/off-line pair is 308 nm vs 353 nm (or 289/299 nm) for stratospheric and tropospheric ozone profiling [106]. These wavelengths have strong ozone absorption cross-sections, allowing sensitive detection. They also benefit from being in a spectral region where solar background can be low (at night) and where Rayleigh scattering provides decent backscatter for range-resolved signals. However, UV/visible light is strongly scattered by air and aerosols, which can limit range (especially in the lower atmosphere under hazy conditions). Moreover, from a HAPS communications standpoint, UV/visible are not typical communications bands (most communications prefer IR or microwave for better penetration and eye safety). Thus, UV wavelengths used for ozone or $SO_2$ DIAL would likely be dedicated purely to sensing. That said, there have been niche proposals for UV communication, but these are low-bandwidth and short-range, not relevant to HAPS high-capacity links. Consequently, UV sensing applications on HAPS, such as ozone detection or aerosol fluorescence, would almost certainly rely on dedicated instrumentation rather than sharing hardware with telecom lasers.

The primary consideration in UV wavelength selection is avoiding absorption by oxygen (the Schumann–Runge bands <200 nm and Hartley band ~250 nm are very opaque). Ozone DIAL picks lines in a "sweet spot" around 280–300 nm or uses a UV laser tuned on/off an ozone absorption line at 308 nm. Similarly, a DIAL for mercury or other metal pollutants might use UV resonance lines around 253 nm, provided suitable laser sources are available. Overall, however, UV sensing from HAPS is likely to remain limited to specialized environmental applications due to strong atmospheric attenuation and minimal overlap with communication wavelength requirements.

### B. Near-Infrared (NIR) 0.7–1.0 μm

This region is important for water vapor sensing. Water vapor has relatively strong overtone absorption bands near 720–730 nm and around 820–940 nm. Indeed, for space and airborne DIAL, the 935 nm band is frequently chosen for water vapor profiling [107,108]. For instance, the HALO aircraft water vapor DIAL transmits four wavelengths around 935 nm to cover the dynamic range of humidity from near-surface to upper troposphere [94]. The 935 nm region is favored because it has moderately strong $H_2O$ lines and lies in a window with sufficient atmospheric transmission (i.e., $CO_2$) to reach high altitudes. Another water band used is ~820 nm (another overtone) for lower troposphere where humidity is high. These NIR wavelengths are accessible with Ti:Sapphire lasers, OPOs pumped by Nd:YAG lasers or diode lasers, and detectors (silicon APDs or PMTs for <1 μm, InGaAs for ~1 μm) are readily available.

For communication, wavelengths in the ~0.8–0.9 μm range are not standard for long-distance optical communications; the atmosphere and fiber attenuation is higher there compared to the 1.3 or 1.55 μm telecom bands. However, short-range optical links or older fiber systems have used 850 nm (e.g., some data centers, multimode fiber). For a HAPS, an 850–900 nm laser could potentially be used for a short air-to-ground link, but it would suffer more from Rayleigh scattering and cloud attenuation than a 1550 nm link. Also, eye safety is more of a concern at 0.8 μm (since the human eye lens focuses it on the retina). Therefore, it is likely that if a HAPS has a 935 nm water vapor DIAL, that laser will not double as a communication link except perhaps at low power for short hops. Instead, the HAPS might carry a separate 1550 nm transmitter for communication. Nevertheless, the NIR (0.7–1.0 μm) region offers some overlap with technologies: Nd:YAG lasers at 946 nm or diode lasers at 905 nm (commonly used in rangefinders) are mature; these could be repurposed for moderate-bandwidth communication or telemetry.

### C. Telecom Infrared (1.3 μm & 1.55 μm bands)

The classic telecom windows in fiber optics are around 1310 nm and 1550 nm, where fiber loss is minimal and components (lasers, EDFAs, photodiodes) are mass-produced. For free-space communications (e.g., laser downlinks from HAPS), 1550 nm is very attractive because atmospheric transmission is quite good (apart from mild water vapor absorption features) and it is eye-safe at high powers (the eye's cornea absorbs most of it, protecting the retina). Conveniently, many target gases have absorption lines in the 1.3–1.6 μm range. Carbon Dioxide ($CO_2$) has a prominent band near 1560 nm (the $2\nu_1+2\nu_2+\nu_3$ band) which has been widely used for DIAL/IPDA. In fact, NASA's ASCENDS program and others demonstrated $CO_2$ DIAL at 1572 nm using fiber laser technology [109]. That wavelength was chosen precisely because it lies in a region of good atmospheric transmission, and coincides with a $CO_2$ absorption line strong enough to measure the column. A big advantage of this selection is that one can leverage telecom-developed components: for example, the $CO_2$ lidar by Abshire et al. used a pulsed fiber amplifier and photon-counting InGaAs/APD detector, all operating near 1.57 μm [109]. This means the laser transceiver can be compact, reliable, and efficient, attributes critical for HAPS payloads. Likewise, methane ($CH_4$) has an absorption band around 1645 nm which the Franco-German MERLIN satellite mission has selected for IPDA measurements [110]. The $CH_4$ line near 1645 nm is actually a cluster of lines that provides suitable absorption without being fully opaque [110]. This wavelength again is close to the telecom C-band. It is feasible to have a laser tune between on-line (~1645.5 nm) and off-line (~1646 nm for example) for methane and use similar InGaAs detectors. For HAPS, the implication is that a single 1.5–1.6 μm laser system could serve dual roles: it can perform $CO_2$/$CH_4$ DIAL sensing and also be modulated for optical communication. The overlap with telecom technology means high data-rate modulators (10–40 Gbps) and optical amplifiers (erbium-doped fiber amplifiers, EDFA) are available at these wavelengths if needed for communication links. Indeed, one could imagine a HAPS with an all-fiber 1572 nm system that during most of the time carries broadband data, but periodically inserts narrow DIAL measurement pulses. Since DIAL pulses are typically low duty-cycle, they would not substantially dent the average communication throughput. Conversely, even while in communication mode, the system could monitor atmospheric absorption on the data channel as demonstrated successfully in [39].

From a sensing perspective, the 1.5 μm region is somewhat of a Goldilocks zone: many gases ($CO_2$, $CH_4$, $H_2O$, CO) have absorption lines there of appropriate strength for a DIAL in ~5–20 km path, and the background sky is also relatively dark (important for daytime operation). From a communication perspective, 1.5 μm is ideal for long-range FSO because it can transmit through atmosphere fairly well (except heavy rain/fog) and is compatible with existing fiber network gateways. Thus, wavelength synergy is maximized in the 1.5 μm band, making it a prime candidate for ISAC. This is evidenced by recent "multi-functional" lidar developments: for instance, Qiang et al. (2024) built an all-fiber DIAL at 1572 nm that can measure $CO_2$ and also use coherent detection, a setup that inherently could be adapted to communication [111]. Thus, considering the maturity of optical systems in the telecom C-band, we focus our analysis on this region, even though it might not be optimum for some species detection, had the sensing task been solely required.

Fig. 8 is organized as a two-tier view of the telecom C-band. Panel (a) overlays the calculated absorbance of two greenhouse gases of interest: $CO_2$ (red) and $N_2O$ (gold), together with the water-vapor background (blue) for a 40 km slant column (equivalent to a 20 km nadir path from the HAPS) at average conditions of 240 K and 0.3 atm. Here, 1,000 ppm and 400 ppm were assumed as column average concentrations for water vapor and $CO_2$, respectively. The HITRAN2020 spectral database was used in these simulations [90]. The plot shows that the strong $CO_2$ feature at 1571.11 nm and the $N_2O$ $R$(24) line at 1564.95 nm sit in windows where water lines are at least two orders of magnitude weaker, providing comfortable signal-to-noise margins for differential retrieval.

Panel (b) presents an exactly analogous overlay for the three species whose spectral signatures are less uniformly present in the atmosphere but of high environmental value when they do appear: $O_3$ (cyan), $CH_4$ (magenta) and $H_2S$ (green). Their absorptions occupy different micro-windows of the same C-band: $O_3$ in a narrow $R$-branch feature at 1564.8 nm, $CH_4$ in a cluster near 1556.5 nm, and $H_2S$ in a broad spectral segment.

Panels (c)–(g) zoom into the individual detection channels selected for deployment: (c) $CO_2$ (1571.11 nm); (d) $N_2O$ (1564.95 nm); (e) $H_2S$ (1567.03 nm); (f) $CH_4$ (1556.5 nm); and (g) $O_3$ (1564.77 nm). In each zoom-in, the plot shows the target gas absorbance at a representative 1,000 ppm (or 400 ppm for $CO_2$) concentration, together with blue and red traces which reproduce the underlying $H_2O$ and $CO_2$ continua. Assuming a minimum detectable absorbance of 0.1%, typical for remote DIAL applications, all five lines exceed this limit by factors of ten to ten-thousand, demonstrating that (with modest optical power and sub-second integration) HAPS can retrieve column-average concentrations in the ppm levels for $CO_2$, $CH_4$, $N_2O$ and $O_3$, and into the tens-of-ppb regime for $H_2S$.

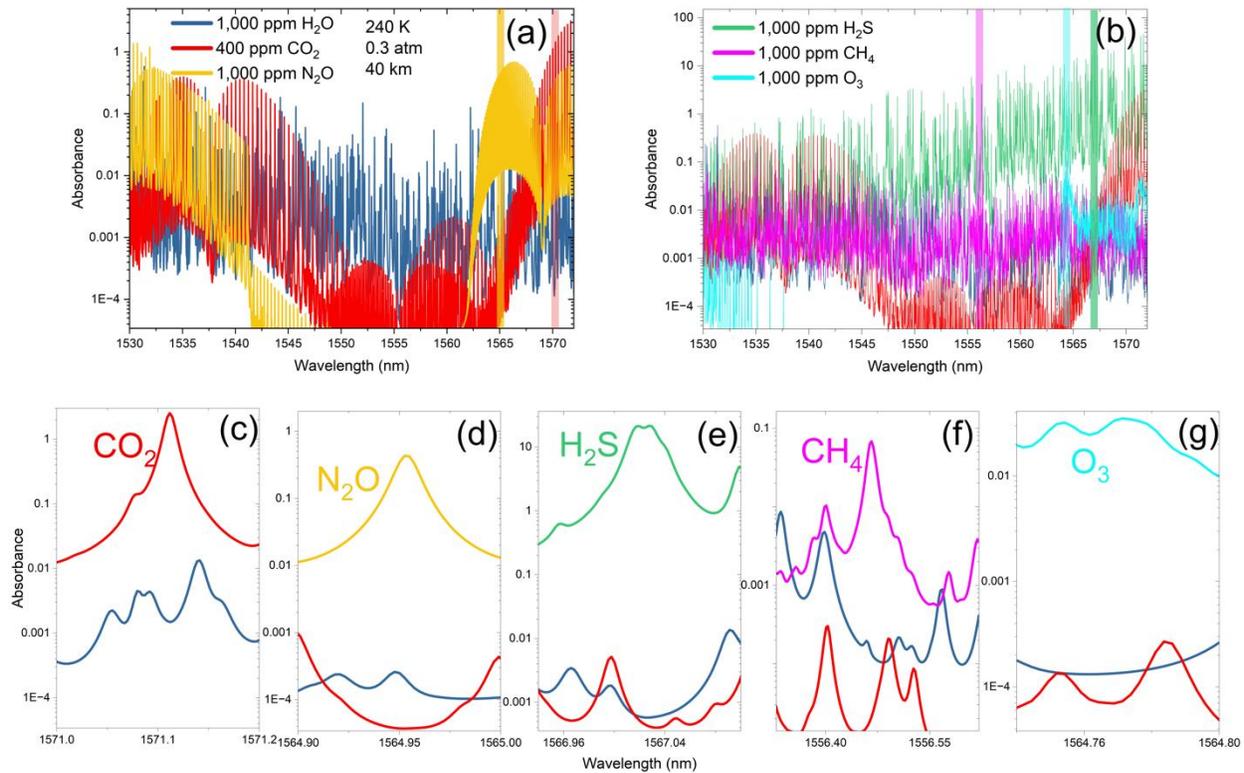

**Fig. 8.** Simulated (HITRAN2020 [90]) absorption spectra over the telecom C-band region for DIAL targeting trace gases from HAPS at 20 km altitude (40-km round path). (a–b) Broadband absorbance spectra for key greenhouse and trace gases. (c–g) Narrowband views of the proposed spectral windows selected.

Because each target wavelength falls inside or adjacent to a 100-GHz ITU grid slot, the Fig. underscores the practicality of interleaving dedicated sensing channels with ordinary DWDM data channels in a single C-band transceiver. A HAPS can therefore act simultaneously as a gigabit relay and as a multi-species spectroscopic sentinel, filling the spatial-resolution gap between surface towers and polar-orbit satellites while using mature telecom hardware.

Based on the spectral simulations, we assess the detection limit of investigated species across the chosen spectral windows while assuming a HAPS altitude of 20 km and a minimum detectable absorbance of 0.1%. Table VI summarizes the results, including the selected absorption features and the rationale for detecting each specific species.

TABLE VI
SUMMARY OF SPECIES-SPECIFIC DETECTION LIMITS FOR DIAL FROM HAPS AT 20 KM ALTITUDE.

| Criterion → / Species ↓ | Selected λ (nm) | Detection limit over 20 km (ppm) | Band/transition quantum number | Importance of high-altitude detection |
|---|---|---|---|---|
| **$CO_2$** | 1571.11 | 1 | $(30012 \leftarrow 00001)$ $R(24)$ | GHG; carbon-flux gap-filler: HAPS bridges the resolution gap between towers and satellites, enabling kilometre-scale flux inversions and continuous validation of spaceborne $CO_2$ missions over target regions. |
| **$N_2O$** | 1564.95 | 2 | $(5000 \leftarrow 0000)$ $R(24)e$ | GHG; regional $N_2O$ budget: a 20 km vantage provides a full-column view below the tropopause, tracking fertiliser & industrial $N_2O$ and its entry into the stratosphere where it drives catalytic ozone loss. |
| **$H_2S$** | 1567.03 | 0.1 | $(111 \leftarrow 000)$ $R(7)$ | Early toxic-gas alert: a 20 km line-of-sight can spot $H_2S$ plumes from sour-gas facilities or volcanic vents hours before they reach populated areas, with alarms relayed via the same HAPS communication link. |
| **$CH_4$** | 1556.47 | 10 | $(0005 \leftarrow 0000)$ $P(3)$ | GHG; super-emitter watchtower: persistent loitering pinpoints episodic $CH_4$ releases (oil-&-gas, landfills, wetlands) that polar-orbit satellites miss, enabling rapid emission mitigation. |
| **$O_3$** | 1564.77 | 50 | $(223 \leftarrow 000)$ cluster of $R$-transitions | Co-located ozone layers: from stratospheric altitude the sensor sees both the overlying ozone column and underlying tropospheric $O_3$, supplying |

| Criterion → / Species ↓ | Selected λ (nm) | Detection limit over 20 km (ppm) | Band/transition quantum number | Importance of high-altitude detection |
|---|---|---|---|---|
| | | | | simultaneous data for ozone-layer recovery studies and air-quality forecasts. |

### D. Mid-Infrared (2–20 μm)

Going further into IR, we encounter very strong fundamental absorption bands of many molecules (e.g., $CO_2$ at 4.3 μm, $CH_4$ at 3.3 μm, $O_3$ at 9–10 μm, $NH_3$ at 10.6 μm, etc.). Mid-IR DIAL can be extremely sensitive because the cross-sections are large, a short path can yield a measurable absorption. For example, a DIAL for $CH_4$ at 3.3 μm could potentially detect very low concentrations over a few km path-length. For HAPS usage, mid-IR DIAL is currently less ideal, but not out of the question. If a particular gas of interest demands it (say HF or hydrocarbons with no good near-IR lines), a HAPS could carry a mid-IR QCL-based sensor.

Free-space optical (FSO) communication and sensing in the mid-infrared (MIR) range have been increasingly validated, with growing evidence supporting their viability for long-distance, open-path applications. Notably, MIR systems experience significantly lower attenuation due to Mie and Rayleigh scattering, making them ideal for deployment in turbid atmospheric environments (e.g., dust, fog, and smoke) that commonly challenge NIR systems [112,113]. Additionally, MIR beams suffer less from atmospheric turbulence and beam wander, which are two major limitations of NIR-based FSO [113,114].

A key enabler of MIR FSO systems is the quantum cascade laser (QCL), which offers high optical power (>3 W) [115–117], wide modulation bandwidths (up to 100 GHz) [118], and robust beam quality [115]. These properties support high-speed data transmission (e.g., > 6 Gbps) and have been successfully demonstrated under simultaneous operation with gas sensing using a hybrid 8-μm QCL system [118]. Our recent experimental study integrated $H_2S$ detection and wireless communication over a shared MIR channel, showing robust performance with BERs below the forward error correction (FEC) limit even at high gas concentrations [39].

Such work demonstrates that MIR systems can operate over long simulated distances (e.g., 20 km) while maintaining sensitivity and signal integrity, which supports the feasibility of their deployment on HAPS for remote sensing and communication. Given the strong molecular absorption features in the MIR (e.g., $H_2S$ at 1234.6 cm$^{-1}$) and the presence of atmospheric transmission windows in the 8–14 μm range (the long-wave midinfrared) [119–124], the MIR domain is especially attractive for DIAL targeting industrial gases.

Yakovlev et al. [86] showed that mid-infrared DIAL is feasible for methane monitoring by deploying an OPO lidar tuned to 3.4157 μm (on-line) and 3.4177 μm (off-line). Operating at 6 mJ and 10 Hz with 100 m range resolution, the ground-based system retrieved a path-averaged $CH_4$ concentration of 2.085 ppm along an 800 m horizontal track; uncertainties (8–13 %) and laboratory cell calibrations matched HITRAN-based expectations, confirming the technique's viability in the 3.3–3.5 μm MIR window. One notable mid-IR wavelength that has seen dual-use consideration is 10.6 μm (the $CO_2$ laser wavelength), which historically was used in long-range laser communications decades ago and is also used in some LIDAR (e.g., older ozone DIALs for $H_2O$ in upper atmosphere).

That said, NIR technology currently remains more favorable for HAPS-based DIAL due to the maturity of its components, including photodetectors with higher detectivity (~$10^{12}$ Jones) and lower noise compared to current MIR detectors (~$10^9$–$10^{11}$ Jones) [125]. Furthermore, existing infrastructure and system-level integration tools are more developed in the NIR range [126–128]. However, as MIR source and detector technologies continue to advance, particularly in the areas of room-temperature QCLs, even at the extreme ends of the MIR [129], and improved MIR photodetectors, MIR-based DIAL systems may soon present a viable alternative to NIR, offering higher specificity, stronger absorption signals, and better performance in adverse environments. Continued development in this field will open the door to high-precision, long-range atmospheric monitoring from HAPS, particularly for industrial and environmental gases (e.g., hydrocarbons and halocarbons) that exhibit strong MIR absorption.

### E. General Considerations

When integrating sensing (DIAL) with communication, interference must be avoided. Typically, the "on-line" sensing wavelength is chosen to be distinct from the communication wavelength, preventing the modulated signal from corrupting the measurement. For instance, if one uses 1572.0 nm for $CO_2$ DIAL on-line, the communication channel might be on the slightly shifted off-line wavelength which is out of the absorption line but close enough to share most optics. Alternatively, fast wavelength tuning or pulsing can separate the functions in time. Some designs might use one laser for both (tune it rapidly between tasks) or use a multi-wavelength laser that can intermix two colors (one for comm, one for sensing). The eye safety and regulatory aspects also influence wavelength: 1550 nm is eye-safe at high power (hence good for free-space links), whereas 1064 nm or 532 nm are not eye-safe at the powers needed for long distance, making 1550 nm even more attractive to cover both LIDAR and communication needs if possible. On the RF side, if a HAPS uses microwave frequencies (e.g., 28 GHz, 60 GHz) for backhaul, those frequencies were likely chosen for spectrum availability and throughput, but interestingly, 60 GHz coincides with an $O_2$ absorption peak [130]. A network might avoid that frequency to reduce losses, or in some cases intentionally use it for $O_2$ mapping; if a HAPS network did operate near an absorption line, it could in theory derive gas information (like using signal attenuation to estimate $O_2$ column, which correlates with air density/pressure).

## X. Optical Payload Constraints for ISAC on HAPS

Integrating both a laser communications terminal and a laser-based gas sensor on a HAPS must reckon with the SWaP limitations of the platform. Each additional instrument and each watt of power draw are precious resources on these high-flying vehicles.

### A. Size & Weight Constraints

HAPS payloads are often measured in only tens of kilograms. For example, a solar-powered fixed-wing HAPS might carry ~20–30 kg total [49], which must include not just the optical systems but also their support electronics, stabilization platforms, cooling systems, and so on. A combined comm-and-sensing payload, if treated as separate subsystems, could quickly encroach on these limits. A telescope for free-space optical communication might, for example, be 10–20 cm in aperture (a few kg with its mount), and a DIAL might require its own telescope or scanning mirror of similar size, plus laser transmitters and detectors. To manage weight, designers look to combine elements: for instance, using one telescope for both functions by inserting beam splitters for different wavelengths, or sharing the inertial stabilizer between the two optical channels. Additionally, components are made of lightweight materials (aluminum or carbon fiber optical benches, lightweight mirrors) to reduce mass. The airborne methane detection systems tested on balloons have been made compact, leveraging fiber-optic components, and those lessons carry to HAPS.

### B. Power Supply and Consumption

Unlike satellites with constant sunlight (in geostationary) or known eclipse cycles, solar HAPS face a daily night cycle requiring energy storage. Power is a major limiting factor: a typical fixed-wing HAPS might generate a few hundred watts in midday and considerably less in mornings/evenings or high latitudes [49]. Airships could generate more power if they are large but 10 kW, for example, would be an upper optimistic Fig. [49], which also has to run propulsion and avionics. Optical communication transmitters (especially if aiming for Gbps rates) often use diode or fiber lasers with output powers of hundreds of milliwatts to a few watts, and wall-plug efficiencies ~20–30%. A continuously operating 1 W optical transmitter might draw ~5 W of electrical power (including modulation electronics). This is quite reasonable. However, a high-power DIAL for gas sensing could require short but powerful pulses; for instance, a DIAL might fire 100 mJ laser pulses in nanoseconds to get enough return signal from the ground. Even at a modest repetition rate, the average optical power could reach tens of watts, and the corresponding peak power requires high-capacitance energy storage and robust laser components. Such a system could easily consume more than 100 W during active operation. Effective power management is therefore critical: the sensor could be duty-cycled, acquiring measurements intermittently (e.g., at a 5% duty cycle) to conserve battery power, and possibly alternating operation with the communication system to minimize peak load overlap. The HAPS power system also has to allocate energy to heating (to keep components at operating temperature in -50°C ambient) and to the basic flight operations (running servos, telemetry, etc.). Future advances in solar cell efficiency and battery energy density will help, but for now, careful budgeting and possibly power-aware scheduling of the two functions is required. For instance, during midday when solar energy is abundant, the HAPS could support both high-speed optical communication and active DIAL sensing. At night, when operating on battery power, it might scale down to low-data-rate communication and infrequent sensing pulses to conserve energy.

### C. Volume and Form Factor

The physical volume available for payloads on HAPS is limited. On a small UAV like *Zephyr*, the payload bay might be only the size of a shoebox or a briefcase. Balloons can carry larger gondolas, but those create more drag (which can shorten flight or require more ballast). The integrated system may need to be extremely compact. This drives innovation in photonics which motivated ISAC in the first place in this context, for instance, using fiber lasers and photonic integrated circuits instead of bulk optical benches. In one demonstration, researchers built a combined LIDAR/FSO system on a silicon photonic chip a few centimeters in size [88], hinting at what future HAPS payloads could employ to save space. Similarly, detector electronics can be compacted using ASICs (application-specific ICs). The overall payload should be aerodynamically packaged, internally housed or enclosed in a streamlined pod under a UAV wing to minimize drag, or in the case of a balloon, mounted within a teardrop-shaped gondola.

### D. Thermal Management

At 20 km, while the air is cold, cooling high-power electronics can paradoxically be difficult because the thin air carries heat away poorly. Lasers and amplifiers generating waste heat might require radiative cooling (large surface area radiators) or heat spreaders. Conversely, some components might need heating; for instance, batteries and optics might need to be kept above a minimum temperature to function properly. All this must be done within the power budget. One strategy is to use the cold ambient as a heat sink at night by radiating excess heat, and in daytime manage temperatures via reflective coatings and perhaps phase-change materials. The design must ensure thermal stability for sensitive laser components, since gas absorption lines shift with temperature; for example, a DFB laser used to target a methane line may require precise temperature control to maintain wavelength accuracy.

### E. Reliability and Autonomy

Because HAPS may stay aloft for months, the payload should be robust against degradation. UV exposure in the stratosphere is intense (ozone layer is below or at those altitudes), so optical coatings must be space-grade to not yellow or peel. Components like moving scanners or mechanical shutters are points of potential failure; designers lean towards solid-state solutions (e.g., acousto-optic or electro-optic beam steering in the future) for longevity. These systems must operate largely autonomously, as human intervention after launch is minimal. Therefore, the integration of communication and sensing should incorporate fault detection and isolation mechanisms, ensuring that a failure in one subsystem does not compromise the other, as well as provisions for remote rebooting, diagnostics, or in-flight recalibration via the communication link.

## XI. Scientometric Analysis and Future Trends

Despite significant technological developments in both domains, the convergence of HAPS and DIAL remains relatively underexplored in the academic literature. To understand the trajectory and maturity of this emerging field, we perform a scientometric analysis covering two decades of publications on HAPS. This analysis quantifies research activity and identifies leading contributors and institutions. Building on these trends, we outline future research opportunities and propose pathways for integrating HAPS-based DIAL systems into climate monitoring, disaster response, and next-generation environmental sensing architectures.

### A. Scientometric Analysis

Using the Scopus database, we carried out a scientometric study covering two decades (2005–2025), employing topic-based searches to retrieve publications related to HAPS. To ensure comprehensive results, a broad query string was used for HAPS including phrases such as "high altitude platform", "high altitude airship", "high altitude long endurance (HALE)", "high altitude pseudo-satellite", and "stratospheric platform/aircraft/airship". The search was done within article titles, abstracts, and keywords. We included peer-reviewed journal articles, conference papers, as well as technical reports where identifiable in Scopus. The final dataset provides a basis for bibliometric indicators including publication counts per year, and leading authors and institutions. All quantitative results shown in Fig. 9 are derived from this dataset.

The publication trajectory depicted in panel (a) reveals the HAPS field remained exploratory until roughly 2011, experienced a consolidation plateau for the next seven years, and then entered a pronounced acceleration after 2019, tripling its annual output to nearly 700 papers by 2024. This surge coincides with the maturation of stratospheric platform prototypes and renewed industrial investment, creating fertile ground for integrating advanced payloads such as DIAL.

Disciplinary profiling in panel (b) shows that more than half of the literature is rooted in engineering and computer science (which are broad umbrellas of fields), underscoring a technology-driven agenda focused on platform design, communications architecture, and control software. Yet sizeable fractions in physics, astronomy, and earth-science categories signal growing scientific demand for high-altitude measurements, precisely the niche where DIAL excels at profiling greenhouse gases, water vapor, and aerosols. As materials-science and energy slices expand, lighter structures and high-density power systems will further ease the integration of eye-safe, multi-wavelength DIAL instruments aboard long-endurance vehicles.

Authorship and affiliation patterns (panels c–e) highlight both concentration and opportunity. A small cohort of prolific authors and institutions, led by Chinese Academy of Sciences, Beihang University, and several key Western universities, accounts for a disproportionate share of output, while China now dominates national contributions. These clusters already have the necessary infrastructure and expertise to develop and deploy airborne and stratospheric DIAL systems. Collaboration between such technology centers and atmospheric science groups could accelerate the demonstration and scaling of HAPS-based DIAL capabilities. The presence of DLR and Carleton University among the top producers suggests an emergent trans-regional network that blends aerospace expertise with optical-wireless communications, an ideal skill set for joint HAPS-DIAL missions.

Overall, the scientometric evidence indicates that the ecosystem is primed for a shift from concept papers to full-scale deployments of DIAL on HAPS. The steep rise in engineering publications supplies the requisite airframes and power budgets, while the steady contributions from physics and earth-science communities ensure a pipeline of environmental applications. Future growth is therefore likely to be measured not only by citation metrics but by operational milestones: multi-month stratospheric flights, continuous 3-D mapping of $CO_2$ and $CH_4$, and real-time assimilation of HAPS-based DIAL profiles into numerical-weather-prediction and climate models.

### B. Future Trends

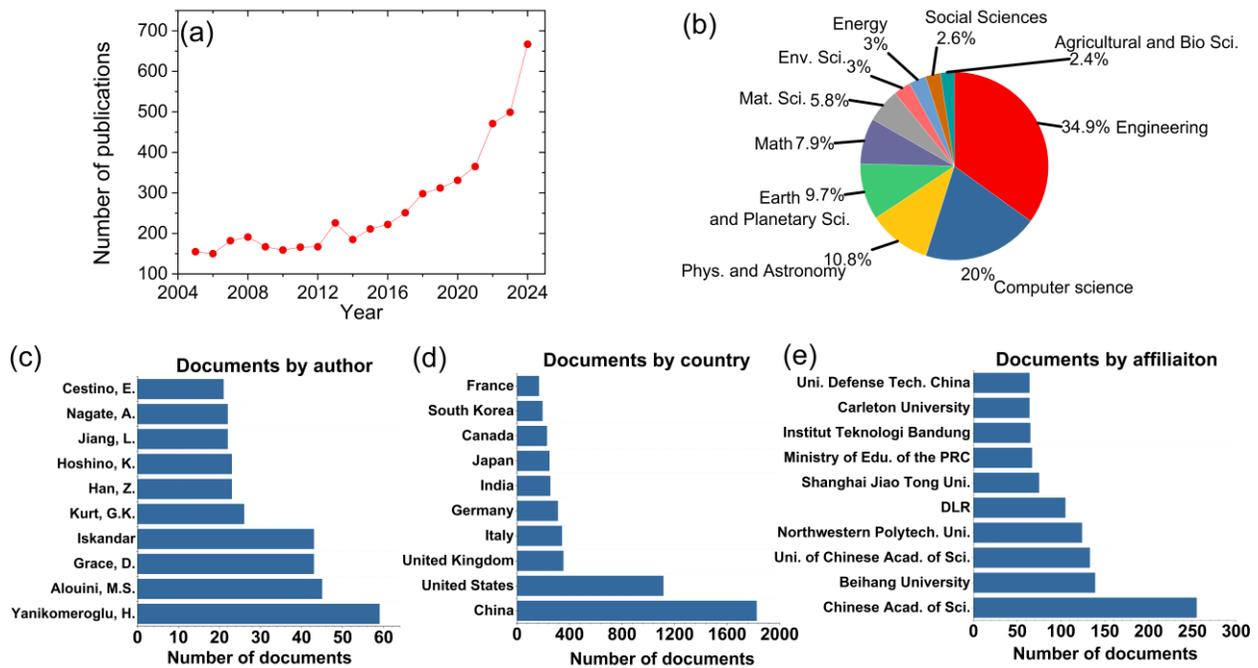

**Fig. 9.** a) Annual number of HAPS-relevant publications from 2005 to 2024. b) Distribution of HAPS documents by subject area; c) top contributing authors; d) countries; and (e) affiliations.

The future likely holds more networked HAPS constellations. Instead of one platform, there might be a fleet of HAPS working in concert, perhaps spanning an entire country or continent in a mesh. These could hand off data and tasks, providing seamless large-area coverage when needed. For sensing, multiple HAPS could even do cooperative measurements, e.g., two DIAL HAPS flying in tandem could do column measurements in stereo to deduce 3D distributions, or one could serve as a transceiver for the other (bistatic lidar configurations for specialized observations). The integration of sensing & communication will likely mature, so future HAPS might routinely carry software-defined payloads that allocate resources to either data relay or sensing based on demand. For instance, at night the telecom demand might dip, so the platform devotes more laser time to environmental scanning. Conversely, in a crisis scenario, the HAPS might prioritize communication but still periodically check environmental variables.

Another trend is miniaturization and automation of DIAL instruments through advances in photonics. The demonstration of an all-fiber $CO_2$ DIAL with integrated single-photon and coherent detection is a hint of things to come [111]. It showed that one compact 1572 nm system could yield multiple data products: $CO_2$ concentration, aerosol backscatter (via the photon-counting channel), and even wind profiles (via the coherent channel) [111]. Such multi-function lidar payloads will greatly enhance HAPS utility, essentially doing the job of multiple instruments at once. By measuring $CO_2$ and wind together, for example, one can directly determine $CO_2$ flux (source/sink strength), which is a breakthrough for carbon cycle science [111]. We may expect future HAPS missions to carry these multi-parameter DIALs for greenhouse gas monitoring projects. NASA's interest in HAPS as part of future Earth observing systems also suggests we may see HAPS included in the global climate observing network, tasked with monitoring areas that are otherwise hard to measure (like the Amazon basin, polar icecap margins, or mid-ocean regions)[48].

In the coming decade, as HAPS technology becomes more mature (with longer flight durations and heavier payload capacities), we might witness operational services such as: "Airborne regional observatories", essentially floating observatory platforms that a government or agency stations over an area to continuously scan the environment. These could be leased or deployed during certain seasons (e.g., fire season, monsoon). The data from HAPS will feed into environmental intelligence systems, likely in real time thanks to the onboard communication links. This ties into the broader vision of "digital earth" and "smart cities," where live data at all scales are assimilated. HAPS uniquely contribute at the mesoscopic scale with both sensing and connectivity, making them a cornerstone of future environmental monitoring infrastructure [95].

In military applications, HAPS can project communication links into denied areas using laser channels that are inherently resistant to jamming, while simultaneously acting as ISR platforms capable of detecting and tracking threats through their spectral signatures or providing target-quality data for precision engagement. All such intelligence can then be transmitted swiftly and securely through the optical network overhead. ISAC on HAPS essentially means these platforms not only see but also speak. They reduce the delay between observation and action by ensuring that large volumes of sensor data do not get bottlenecked (as can happen with satellites that store data onboard until a ground pass).

To realize this future, interdisciplinary efforts are required: aerospace engineers, optical scientists, atmospheric experts, communications network designers, and AI specialists must collaborate. Investment in HAPS testbeds and prototypes will be needed to iron out practical issues (from optical alignment to regulatory concerns about using lasers in airspace). There are also regulatory and spectrum aspects; optical frequencies are unlicensed, but ensuring safe laser use (eyesafe operations, coordination with aviation authorities) will be important as HAPS deployments increase. The trajectory is clear, however. HAPS combine the flexibility of aircraft with the endurance of satellites, carving out a unique niche. By equipping them with versatile optical payloads that can both connect and detect, we stand to gain an unprecedented capability for real-time understanding and communication across the planet. The coming decade will likely see the first operational systems that embody this vision, and their success could herald a new layer of infrastructure high above us, an "optical layer" of HAPS nodes seamlessly integrating into our communication networks and our sensor webs, ultimately helping to bridge the digital divide, respond to disasters faster, and keep watch over Earth's atmosphere with a fidelity never before achieved.

The coming years hold many opportunities to enhance the combined optical communication and sensing capabilities of HAPS. Below, we outline several key directions for research and development that will shape the future of these multi-functional HAPS.

*1) Advanced Optical Components and Integration:* Continued innovation in photonics will drive HAPS capabilities. Developing multi-wavelength laser systems that can perform both communication and spectroscopy is one avenue; for instance, a tunable laser that can rapidly switch from a data-transmission wavelength to a gas-sensing wavelength (or even use a comb of wavelengths) would simplify payloads. Likewise, large yet lightweight telescopes or diffractive optical elements could improve link budgets and sensing range. There is interest in photonic integrated circuits (PICs) that integrate lasers, modulators, and detectors on a chip, drastically reducing size and weight. A future HAPS might have a PIC as the heart of its optical payload, handling multiple channels of communication and sensing with minor modifications in software. Materials research for coatings that withstand UV and atomic oxygen at altitude will also extend the life of optical components. Ultimately, a tightly integrated optical system (where perhaps a single box handles both downlink and DIAL) will reduce SWaP and cost per platform.

*2) Adaptive Optics and Beam Control:* To push data rates higher (towards tens of Gbps or more) and to ensure reliable sensing at long distances, adaptive optics (AO) will likely be adopted. AO systems can compensate for dynamic aberrations in the optical path, such as residual turbulence or platform jitter. Compact deformable mirrors and wavefront sensors could be deployed on HAPS optical terminals to sharpen the beam focus and counteract atmospheric distortion in real time (much like astronomy telescopes negate twinkling of stars). For communication, this could mean maintaining error-free links even during periods of turbulence. For DIAL, it could improve the signal return by keeping the beam tightly on target and maximizing coupling of return photons into the receiver. Research into fast beam steering and tracking will also continue; for instance, optical phased arrays (which steer beams electronically with no moving parts) could allow near-instantaneous pointing changes, enabling a HAPS to handle multiple ground stations or scan many points of interest very rapidly. Such agility would enhance both network throughput (servicing multiple links sequentially) and sensing (mapping a larger area faster).

*3) Energy Management and Power Innovations:* Power will remain a gating factor. R&D in high-efficiency solar cells, including multi-junction cells that can harvest a broader spectrum of sunlight, directly translates to more wattage available on HAPS. Similarly, next-generation batteries or supercapacitors with higher energy densities will prolong nighttime operations. There is also the concept of power beaming: using laser or microwave beams from the ground to send power to the HAPS (received by photovoltaic or rectifying antennas). If realized, this could significantly augment a HAPS's power budget, allowing continuous high-power laser operation even at night. For ISAC specifically, more efficient lasers (like diode-pumped fiber lasers with >50% efficiency) and low-power-consumption signal processing (using ASICs or FPGAs optimized for the task) will help stretch limited energy. Thermal control research, such as passive radiators that can reject heat even in thin air, or phase-change materials to buffer thermal loads, will support these high-power optical systems by managing waste heat without heavy active cooling.

*4) AI for Data Processing and Autonomous Operations:* With the dual roles of HAPS, onboard artificial intelligence (AI) and machine learning will be valuable in handling data and optimizing operations. For communications, AI could manage the network, selecting the best modulation and coding schemes on the fly to adapt to atmospheric conditions (akin to cognitive radio but for optical links). For sensing, machine learning can be used onboard to identify features in the data, for example, spotting an anomalous spike in gas concentration or recognizing the signature of a particular chemical agent as demonstrated in [131–134]. Instead of downlinking raw sensor data, which could be prohibitively large in volume, the HAPS could transmit high-level, processed information such as "methane leak detected at these coordinates with this flux." This approach dramatically reduces data size and ensures more efficient use of the available communication bandwidth. This local processing is especially important if the communication link capacity at any point is reduced, for example, due to clouds or switching to a backup RF link. Moreover, AI can assist in sensor fusion: correlating inputs from multiple sensors (visual, infrared, gas, etc.) to draw a more complete picture, which can then be communicated. Autonomous decision-making algorithms will also allow a HAPS to re-task itself; for instance, if the gas sensor detects something noteworthy, the system could autonomously decide to enter a high-data-rate communication mode to immediately relay that information to ground, or conversely, if the communication link is idle, to spend more time on scientific scanning. Such autonomy will be crucial as HAPS deployments scale up (one cannot have a human manually commanding each behavior for dozens of HAPS in real time).

*5) Robustness and Weather Mitigation:* Overcoming the vulnerability of optical systems, e.g., cloud blockage, is an area of active exploration. Solutions include deploying multiple HAPS in a network so that if one's downlink is clouded, another can relay its data via an inter-HAPS optical link to a clear region for downlink [57]. This concept of a mesh network in the sky will require research into optimal network topologies, dynamic routing, and handoff protocols for optical links. Ensuring interoperability between different platforms (e.g., a balloon and a fixed-wing HAPS passing data) will also be important. On the laser side, there is interest in using different wavelengths to mitigate partial obscurations; for instance, shortwave infrared or even terahertz waves might penetrate thin clouds better for communication (though with a trade-off in rate) or using DIAL wavelengths that are less absorbed by aerosols for certain measurements. Having multi-spectral or multi-modal communications/sensing (including perhaps a hybrid optical/RF approach) could make the overall system to be more all-weather. Research into predictive modeling is relevant as well: using weather forecast data to predict where clouds will be and maneuver or schedule HAPS operations accordingly (for example, cache data during a cloudy spell and dump it when a predicted clear window opens).

*6) Field Trials and Demonstrations:* As a forward-looking endeavor, one of the most convincing steps will be integrated field demonstrations. Past experiments have individually shown the feasibility of components: the STROPEX trial showed multi-Gbps optical downlink from 22 km [57], and NASA's HALO airborne lidar showcased high-altitude methane detection [135]. The logical next step is a pilot project that combines these ISAC systems, for instance, flying a stratospheric balloon or UAV with both a high-rate optical terminal and a DIAL, to test how they perform together. Such a demonstration would yield invaluable insights into co-operation challenges, e.g., whether the communication link performance is affected during DIAL operation, how task scheduling should be optimized, and how much benefit real-time data delivery provides. Within the next decade, we envision prototypes where a HAPS continuously monitors greenhouse gases over a region and streams the processed data directly to users on the ground. International collaborations, such as those fostered by the HAPS Alliance or space agencies, could lead joint demonstration campaigns showcasing environmental applications; for instance, a HAPS mapping a methane hotspot with instantaneous data relay as a compelling proof of concept for climate monitoring efforts.

## XII. Conclusions

This tutorial-cum-survey has presented a comprehensive overview of High-Altitude Platform Stations (HAPS) as enablers of next-generation optical communication and atmospheric sensing. By unifying free-space optical (FSO) backhaul and laser-based environmental monitoring under a shared hardware and spectral framework, the study demonstrated the feasibility and promise of Integrated Sensing and Communication (ISAC) architectures on stratospheric platforms. Through a detailed examination of HAPS configurations, link geometries, and optical propagation characteristics, the paper established the physical basis for dual-use operation. Theoretical and simulation analyses further identified viable spectral windows in the telecom C-band (1.53–1.57 μm) that support both high-throughput data transfer and high-sensitivity DIAL sensing of key atmospheric species.

The findings underscore that HAPS, operating between 17 and 22 km, occupy an optimal regime for co-deploying optical communication and environmental monitoring functions. Their persistence, low latency, and large coverage footprint enable real-time data relay, while their proximity to the Earth's surface facilitates precise and responsive sensing of greenhouse gases and pollutants. The synergy of these capabilities positions HAPS as a strategic bridge between satellite and terrestrial systems within future non-terrestrial networks (NTNs).

The scientometric analysis revealed exponential growth in HAPS-related research, marking a transition from conceptual studies to near-operational maturity. As optical, photonic, and power technologies continue to advance, ISAC-enabled HAPS are expected to become integral components of 6G infrastructure and global Earth observation frameworks. Future research should prioritize field demonstrations that integrate FSO and DIAL payloads, mid-infrared system extensions, and intelligent mission management through AI-driven autonomy.

## References

[1] Silva BN, Khan M, Han K. Towards sustainable smart cities: A review of trends, architectures, components, and open challenges in smart cities. Sustain Cities Soc 2018;38:697–713. https://doi.org/10.1016/J.SCS.2018.01.053.

[2] Manie YC, Yao CK, Yeh TY, Teng YC, Peng PC. Laser-Based Optical Wireless Communications for Internet of Things (IoT) Application. IEEE Internet Things J 2022;9:24466–76. https://doi.org/10.1109/JIOT.2022.3190619.

[3] Abbasi O, Yanikomeroglu H. UxNB-Enabled Cell-Free Massive MIMO With HAPS-Assisted Sub-THz Backhauling. IEEE Trans Veh Technol 2024;73:6937–53. https://doi.org/10.1109/TVT.2023.3347140.

[4] Aijaz A, Bean K. Millimeter Wave Backhaul for Stratospheric HAPs: Performance of the Bent-Pipe Architecture. 2023 IEEE Virtual Conference on Communications, VCC 2023 2023:183–7. https://doi.org/10.1109/VCC60689.2023.10475069.

[5] Alfattani S, Jaafar W, Yanikomeroglu H, Yongacoglu A. Multimode High-Altitude Platform Stations for Next-Generation Wireless Networks: Selection Mechanism, Benefits, and Potential Challenges. IEEE Vehicular Technology Magazine 2023;18:20–8. https://doi.org/10.1109/MVT.2023.3289630.

[6] Alam MS, Kurt GK, Yanikomeroglu H, Zhu P, Dao ND. High altitude platform station based super macro base station constellations. IEEE Communications Magazine 2021;59:103–9. https://doi.org/10.1109/MCOM.001.2000542.

[7] Abbasi O, Yadav A, Yanikomeroglu H, Dão ND, Senarath G, Zhu P. HAPS for 6G Networks: Potential Use Cases, Open Challenges, and Possible Solutions. IEEE Wirel Commun 2024;31:324–31. https://doi.org/10.1109/MWC.012.2200365.

[8] Lou Z, Youcef Belmekki BE, Alouini MS. HAPS in the Non-Terrestrial Network Nexus: Prospective Architectures and Performance Insights. IEEE Wirel Commun 2023;30:52–8. https://doi.org/10.1109/MWC.004.2300198.

[9] Alghamdi R, Dahrouj H, Al-Naffouri TY, Alouini MS. Equitable 6G Access Service Via Cloud-Enabled HAPS for Optimizing Hybrid Air-Ground Networks. IEEE Transactions on Communications 2024;72:2959–73. https://doi.org/10.1109/TCOMM.2023.3348842.

[10] Belmekki BEY, Aljohani AJ, Althubaity SA, Harthi A Al, Bean K, Aijaz A, et al. Cellular Network From the Sky: Toward People-Centered Smart Communities. IEEE Open Journal of the Communications Society 2024;5:1916–36. https://doi.org/10.1109/OJCOMS.2024.3378297.

[11] Xi Q, He C, Jiang L, Tian J, Shen Y. Capacity analysis of massive MIMO on high altitude platforms. Proceedings - IEEE Global Communications Conference, GLOBECOM 2016. https://doi.org/10.1109/GLOCOM.2016.7841647.

[12] Xing Y, Hsieh F, Ghosh A, Rappaport TS. High Altitude Platform Stations (HAPS): Architecture and System Performance. IEEE Vehicular Technology Conference 2021;2021-April. https://doi.org/10.1109/VTC2021-SPRING51267.2021.9448899.

[13] Yuniarti D. Regulatory challenges of broadband communication services from High Altitude Platforms (HAPs). 2018 International Conference on Information and Communications Technology, ICOIACT 2018 2018;2018-January:919–22. https://doi.org/10.1109/ICOIACT.2018.8350752.

[14] Abderrahim W, Amin O, Shihada B. Data Center-Enabled High Altitude Platforms: A Green Computing Alternative. IEEE Trans Mob Comput 2024;23:6149–62. https://doi.org/10.1109/TMC.2023.3316204.

[15] Jirousek M, Peichl M, Anger S, Dill S, Limbach M. The DLR High Altitude Platform Synthetic Aperture Radar Instrument HAPSAR | VDE Conference Publication | IEEE Xplore. EUSAR 2024; 15th European Conference on Synthetic Aperture Radar, 2024.

[16] Jirousek M, Peichl M, Anger S, Dill S, Engel M. Design of a Synthetic Aperture Radar Instrument for a High-Altitude Platform. IEEE International Symposium on Geoscience and Remote Sensing (IGARSS), Institute of Electrical and Electronics Engineers (IEEE); 2023, p. 2045–8. https://doi.org/10.1109/IGARSS52108.2023.10283363.

[17] Jirousek M, Peichl M, Anger S, Engel M, Dill S, Scheiber R, et al. Synthetic Aperture Radar Design for a High-Altitude Platform. VDE Conference Publication, 2022.

[18] Huang B, Ahmed S, Alouini MS. Design of Frequency Index Modulated Waveforms for Integrated SAR and Communication on High-Altitude Platforms (HAPs). IEEE Transactions on Communications 2025. https://doi.org/10.1109/TCOMM.2025.3591161.

[19] Brauchle J, Gstaiger V, Azimi S, Muehlhaus M, Nikodem F, Dern P, et al. MACS-HAP: Design and Image Processing Features of the DLR HAP Camera System. International Geoscience and Remote Sensing Symposium (IGARSS) 2023;2023-July:4804–7. https://doi.org/10.1109/IGARSS52108.2023.10282470.

[20] Shah Alamgir M, Kelley B. AI Fusion of Vision and Management for 6G Millimeter-Wave Aerial Networks. IEEE Access 2025;13:152729–47. https://doi.org/10.1109/ACCESS.2025.3604288.

[21] Herman DI, Weerasekara C, Hutcherson LC, Giorgetta FR, Cossel KC, Waxman EM, et al. Precise multispecies agricultural gas flux determined using broadband point-dual-comb spectroscopy. Sci Adv 2021;7:9765–96. https://doi.org/10.1126/SCIADV.ABE9765/SUPPL_FILE/ABE9765_SM.PDF.

[22] Rieker GB, Giorgetta FR, Swann WC, Kofler J, Zolot AM, Sinclair LC, et al. Frequency-comb-based remote sensing of greenhouse gases over kilometer air paths. Optica 2014;1:290. https://doi.org/10.1364/optica.1.000290.

[23] Farooq A, Alquaity ABS, Raza M, Nasir EF, Yao S, Ren W. Laser sensors for energy systems and process industries: Perspectives and directions. Prog Energy Combust Sci 2022;91:100997. https://doi.org/10.1016/J.PECS.2022.100997.

[24] Muraviev A V., Smolski VO, Loparo ZE, Vodopyanov KL. Massively parallel sensing of trace molecules and their isotopologues with broadband subharmonic mid-infrared frequency combs. Nat Photonics 2018;12:209–14. https://doi.org/10.1038/s41566-018-0135-2.

[25] Le Bris K. LABORATORY MID-INFRARED ABSORPTION CROSS-SECTION SPECTRA OF LARGE VOLATILE MOLECULES FOR ATMOSPHERIC REMOTE SENSING. Physics in Canada 2017;73.

[26] Dobler JT, Zaccheo TS, Pernini TG, Blume N, Broquet G, Vogel F, et al. Demonstration of spatial greenhouse gas mapping using laser absorption spectrometers on local scales. J Appl Remote Sens 2017;11:14002. https://doi.org/10.1117/1.JRS.11.014002.

[27] Mei L, Brydegaard M. Continuous-wave differential absorption lidar. Laser Photon Rev 2015;9:629–36. https://doi.org/10.1002/LPOR.201400419.

[28] Koch GJ, Barnes BW, Petros M, Beyon JY, Amzajerdian F, Yu J, et al. Coherent differential absorption lidar measurements of $CO_2$. Applied Optics, Vol 43, Issue 26, Pp 5092-5099 2004;43:5092–9. https://doi.org/10.1364/AO.43.005092.

[29] Browell E V., Wilkerson TD, Mcilrath TJ. Water vapor differential absorption lidar development and evaluation. Applied Optics, Vol 18, Issue 20, Pp 3474-3483 1979;18:3474–83. https://doi.org/10.1364/AO.18.003474.

[30] Browell E V., Ismail S, Grant WB. Differential absorption lidar (DIAL) measurements from air and space. Appl Phys B 1998;67:399–410. https://doi.org/10.1007/S003400005523/METRICS.

[31] Spinelli N, Amodeo A, Girolamo P Di, Ambrico PF. Sensitivity analysis of differential absorption lidar measurements in the mid-infrared region.

[32] Zanrzottera E, Zielinski WL. Differential Absorption Lidar Techniques in the Determination of Trace Pollutants and Physical Parameters of the Atmosphere. Crit Rev Anal Chem 1990;21:279–319. https://doi.org/10.1080/10408349008051632.

[33] Javed S, Alouini MS. System Design and Parameter Optimization for Remote Coverage from NOMA-based High-Altitude Platform Stations (HAPS). IEEE Trans Wirel Commun 2024. https://doi.org/10.1109/TWC.2024.3508872.

[34] Fettes E, Madoery PG, Yanikomeroglu H, Kurt GK, Bellinger C, Martel S, et al. Next-Generation Satellite IoT Networks: A HAPS-Enabled Solution to Enhance Optical Data Transfer 2024.

[35] Pedone M, Aiuppa A, Giudice G, Grassa F, Cardellini C, Chiodini G, et al. Volcanic $CO_2$ flux measurement at Campi Flegrei by tunable diode laser absorption spectroscopy. Bull Volcanol 2014;76:1–13. https://doi.org/10.1007/S00445-014-0812-Z/FIGURES/9.

[36] Ata Y, Alouini MS. HAPS Based FSO Links Performance Analysis and Improvement with Adaptive Optics Correction. IEEE Trans Wirel Commun 2023;22:4916–29. https://doi.org/10.1109/TWC.2022.3230737.

[37] He H, Jiang L, Pan Y, Yi A, Zou X, Pan W, et al. Integrated sensing and communication in an optical fibre. Light: Science & Applications 2023 12:1 2023;12:1–14. https://doi.org/10.1038/s41377-022-01067-1.

[38] Lu S, Liu F, Li Y, Zhang K, Huang H, Zou J, et al. Integrated Sensing and Communications: Recent Advances and Ten Open Challenges. IEEE Internet Things J 2024;11:19094–120. https://doi.org/10.1109/JIOT.2024.3361173.

[39] Elkhazraji A, Elkhazraji A, Sait M, Farooq A. Integrated optical gas sensing and wireless communication in the mid-infrared. Applied Optics, Vol 64, Issue 10, Pp D114-D121 2025;64:D114–21. https://doi.org/10.1364/AO.559367.

[40] D'Oliveira FA, De Melo FCL, Devezas TC. High-Altitude Platforms — Present Situation and Technology Trends. Journal of Aerospace Technology and Management 2016;8:249–62. https://doi.org/10.5028/JATM.V8I3.699.

[41] Qiu J, Grace D, Ding G, Zakaria MD, Wu Q. Air-Ground Heterogeneous Networks for 5G and beyond via Integrating High and Low Altitude Platforms. IEEE Wirel Commun 2019;26:140–8. https://doi.org/10.1109/MWC.0001.1800575.

[42] Arum SC, Grace D, Mitchell PD. A review of wireless communication using high-altitude platforms for extended coverage and capacity. Comput Commun 2020;157:232–56. https://doi.org/10.1016/J.COMCOM.2020.04.020.

[43] Kurt GK, Khoshkholgh MG, Alfattani S, Ibrahim A, Darwish TSJ, Alam MS, et al. A Vision and Framework for the High Altitude Platform Station (HAPS) Networks of the Future. IEEE Communications Surveys and Tutorials 2021;23:729–79. https://doi.org/10.1109/COMST.2021.3066905.

[44] Mershad K, Dahrouj H, Sarieddeen H, Shihada B, Al-Naffouri T, Alouini MS. Cloud-Enabled High-Altitude Platform Systems: Challenges and Opportunities. Frontiers in Communications and Networks 2021;2:716265. https://doi.org/10.3389/FRCMN.2021.716265/BIBTEX.

[45] Services enabled by High Altitude Pseudo Satellites (HAPS) complemented by satellites n.d. https://business.esa.int/funding/invitation-to-tender/services-enabled-high-altitude-pseudo-satellites-haps-complemented-satellites (accessed June 26, 2025).

[46] Thales Alenia Space's Stratobus stratospheric airship passes a new development milestone | Thales Group n.d. https://www.thalesgroup.com/en/news-centre/press-releases/thales-alenia-spaces-stratobus-stratospheric-airship-passes-new (accessed October 15, 2025).

[47] Stratospheric Platforms for a Better Planet - Sceye n.d. https://sceye.com/ (accessed October 15, 2025).

[48] Fladeland MM. Recent Advances in High Altitude Pseudosatellites (HAPS) and Potential Roles in Future Earth Observing Systems 2019.

[49] High Altitude Platform Systems: Towers in the Skies (Version 2.0) 2022. https://www.gsma.com/solutions-and-impact/technologies/networks/gsma_resources/high-altitude-platform-systems-towers-in-the-skies-version-2-0/ (accessed June 26, 2025).

[50] WORLD VIEW. Pioneers of the stratosphere n.d. https://www.worldview.space/remote-sensing (accessed June 26, 2025).

[51] Grest H. High-Altitude Platform Systems Alternative, Supplement, or Competition to Satellites? 2022. https://www.japcc.org/articles/high-altitude-platform-systems/ (accessed June 26, 2025).

[52] HySpex. HySpex payloads successfully demonstrate capabilities in groundbreaking diurnal stratospheric flight n.d. https://www.hyspex.com/about/news/press-releases/press-release-sceye-roswell/ (accessed June 26, 2025).

[53] Dargahi A, Safari M, Safi H, Cheng J. Analytical Channel Model and Link Design Optimization for Ground-to-HAP Free-Space Optical Communications. Journal of Lightwave Technology, Vol 38, Issue 18, Pp 5036-5047 2020;38:5036–47.

[54] Sharma M, Chadha D, Chandra V. High-Altitude Platform for Free-Space Optical Communication: Performance Evaluation and Reliability Analysis. Journal of Optical Communications and Networking, Vol 8, Issue 8, Pp 600-609 2016;8:600–9. https://doi.org/10.1364/JOCN.8.000600.

[55] Bithas PS. Distributed HAPS-assisted communications in FSO/RF space-air-ground integrated networks. Opt Quantum Electron 2024;56:1–16. https://doi.org/10.1007/S11082-024-06918-2/FIGURES/4.

[56] QinetiQ. Free Space Optical Communications n.d. https://www.qinetiq.com/en/what-we-do/services-and-products/free-space-optical-communications (accessed June 26, 2025).

[57] Knapek M, Horwath J, Perlot N, Wilkerson B. The DLR ground station in the optical payload experiment (STROPEX): results of the atmospheric measurement instruments. Https://DoiOrg/101117/12679004 2006;6304:494–504. https://doi.org/10.1117/12.679004.

[58] Wilquet V, Fedorova A, Montmessin F, Drummond R, Mahieux A, Vandaele AC, et al. Preliminary characterization of the upper haze by SPICAV/SOIR solar occultation in UV to mid-IR onboard Venus Express. J Geophys Res Planets 2009;114:0–42. https://doi.org/10.1029/2008JE003186.

[59] Nazaryan H, McCormick MP, Russell JM. New studies of SAGE II and HALOE ozone profile and long-term change comparisons. Journal of Geophysical Research D: Atmospheres 2005;110:1–13. https://doi.org/10.1029/2005JD005425;REQUESTEDJOURNAL:JOURNAL:21562202D;PAGE:STRING:ARTICLE/CHAPTER.

[60] Lee C, Martin R V., Van Donkelaar A, Richter A, Burrows JP, Kim YJ. Remote Sensing of tropospheric trace gases (NO2 and SO 2) from SCIAMACHY. Atmospheric and Biological Environmental Monitoring 2009:63–72. https://doi.org/10.1007/978-1-4020-9674-7_5.

[61] Yang T, Si F, Zhou H, Zhao M, Lin F, Zhu L. Preflight Evaluation of the Environmental Trace Gases Monitoring Instrument with Nadir and Limb Modes (EMI-NL) Based on Measurements of Standard NO2 Sample Gas. Remote Sensing 2022, Vol 14, Page 5886 2022;14:5886. https://doi.org/10.3390/RS14225886.

[62] Michel DT, Augère B, Boulant T, Cézard N, Dolfi-Bouteyre A, Durécu A, et al. Heterodyne and Direct Detection Wind Lidar Developed at ONERA. Springer Aerospace Technology 2024;Part F2592:227–38. https://doi.org/10.1007/978-3-031-53618-2_20/FIGURES/5.

[63] Hoffmann A, Macleod NA, Huebner M, Weidmann D. Thermal infrared laser heterodyne spectroradiometry for solar occultation atmospheric CO2 measurements. Atmos Meas Tech 2016;9:5975–96. https://doi.org/10.5194/AMT-9-5975-2016,.

[64] Sugimoto N, Chan KP, Killinger DK. Optimal heterodyne detector array size for 1-μm coherent lidar propagation through atmospheric turbulence. Applied Optics, Vol 30, Issue 18, Pp 2609-2616 1991;30:2609–16. https://doi.org/10.1364/AO.30.002609.

[65] Bolek A, Heimann M, Göckede M. UAV-based in situ measurements of CO2 and CH4 fluxes over complex natural ecosystems. Atmos Meas Tech 2024;17:5619–36. https://doi.org/10.5194/AMT-17-5619-2024,.

[66] Satellite Discovery of Anomalously Large Methane Point Sources From Oil/Gas Production | NASA Airborne Science Program n.d. https://airbornescience.nasa.gov/content/Satellite_Discovery_of_Anomalously_Large_Methane_Point_Sources_From_Oil_Gas_Production (accessed June 26, 2025).

[67] Ismail S, Browell E V. LIDAR | Differential Absorption Lidar. Encyclopedia of Atmospheric Sciences: Second Edition 2015:277–88. https://doi.org/10.1016/B978-0-12-382225-3.00204-8.

[68] Ozawa K, Koga N, Sugimoto N, Aoki T, Itabe T, Kunimori H, et al. Optical characteristics of the Retroreflector in Space (RIS) on the ADEOS satellite. Https://DoiOrg/101117/12295657 1997;3218:2–9. https://doi.org/10.1117/12.295657.

[69] Koga N, Sugimoto N, Ozawa K, Saito Y, Nomura A, Minato A, et al. Laser long-path absorption experiment using the Retroreflector in Space (RIS) on the ADEOS satellite. Https://DoiOrg/101117/12295643 1997;3218:10–8. https://doi.org/10.1117/12.295643.

[70] Aoki T, Sugimoto N, Mizutani K, Itabe T. System Accuracy of Atmospheric Observation by RIS (Retroreflector in Space). Optical Remote Sensing of the Atmosphere (1997), Paper OMD10 1997:OMD.10. https://doi.org/10.1364/ORSA.1997.OMD.10.

[71] Sugimoto N. Earth-satellite-earth Laser Long-path Absorption Experiments Using The Retroreflector In Space n.d.

[72] Heaps WS, Hudson RD, McGee TJ, Caudill LO. Stratospheric ozone and hydroxyl radical measurements by balloon-borne lidar. Applied Optics, Vol 21, Issue 12, Pp 2265-2274 1982;21:2265–74. https://doi.org/10.1364/AO.21.002265.

[73] Mcgee TJ. Differential absorption lidar (DIAL) measurements of stratospheric ozone. Network for the Detection of Stratospheric Change 1986.

[74] Nehrir AR, Barton-Grimley RA, Collins JE, Collister B, Crosbie E, DiGangi JP, et al. City-Scale Methane Retrievals from the HALO lidar During the 2023 STAQS Campaign. AGUFM 2023;2023:A51U-2252.

[75] Refaat TF, Ismail S, Nehrir AR, Hair JW, Crawford JH, Leifer I, et al. Performance evaluation of a 1.6-μm methane DIAL system from ground, aircraft and UAV platforms. Optics Express, Vol 21, Issue 25, Pp 30415-30432 2013;21:30415–32. https://doi.org/10.1364/OE.21.030415.

[76] MERLIN | CNES n.d. https://cnes.fr/en/projects/merlin (accessed June 26, 2025).

[77] Wang R, Xie P, Xu J, Li A, Sun Y. Observation of CO2 Regional Distribution Using an Airborne Infrared Remote Sensing Spectrometer (Air-IRSS) in the North China Plain. Remote Sensing 2019, Vol 11, Page 123 2019;11:123. https://doi.org/10.3390/RS11020123.

[78] Lin B, Harrison FW, Obland MD, Browell E V., Campbell J, Nehrir AR, et al. Atmospheric CO2 column measurements in cloudy conditions using intensity-modulated continuous-wave lidar at 1.57 micron. Optics Express, Vol 23, Issue 11, Pp A582-A593 2015;23:A582–93. https://doi.org/10.1364/OE.23.00A582.

[79] Yu J, Refaat TF, Remus R, Singh UN, Petros M, Ismail S. Double-pulse 2-BC;m integrated path differential absorption lidar airborne validation for atmospheric carbon dioxide measurement. Applied Optics, Vol 55, Issue 15, Pp 4232-4246 2016;55:4232–46. https://doi.org/10.1364/AO.55.004232.

[80] Wang S, Ke J, Chen S, Zheng Z, Cheng C, Tong B, et al. Performance Evaluation of Spaceborne Integrated Path Differential Absorption Lidar for Carbon Dioxide Detection at 1572 nm. Remote Sensing 2020, Vol 12, Page 2570 2020;12:2570. https://doi.org/10.3390/RS12162570.

[81] Ehret G, Bousquet P, Pierangelo C, Alpers M, Millet B, Abshire JB, et al. MERLIN: A French-German Space Lidar Mission Dedicated to Atmospheric Methane. Remote Sensing 2017, Vol 9, Page 1052 2017;9:1052. https://doi.org/10.3390/RS9101052.

[82] Sugimoto N, Koga N, Matsui I, Sasano Y, Minato A, Ozawa K, et al. Earth-satellite-Earth laser long-path absorption experiment using the Retroreflector in Space (RIS) on the Advanced Earth Observing Satellite (ADEOS). Journal of Optics A: Pure and Applied Optics 1999;1:201. https://doi.org/10.1088/1464-4258/1/2/015.

[83] Chen W, Ke J, Zhang Y, Cheng C, Liu D, Zheng Z, et al. Performance estimation of space-borne high-spectral-resolution lidar for cloud and aerosol optical properties at 532 nm. Optics Express, Vol 27, Issue 8, Pp A481-A494 2019;27:A481–94. https://doi.org/10.1364/OE.27.00A481.

[84] Zhou DK, Larar AM, Liu X, Noe AM, Diskin GS, Soja AJ, et al. Wildfire-Induced CO Plume Observations from NAST-I during the FIREX-AQ Field Campaign. IEEE J Sel Top Appl Earth Obs Remote Sens 2021;14:2901–10. https://doi.org/10.1109/JSTARS.2021.3059855.

[85] Stroud JR, Wagner GA, Plusquellic DF. Multi-Frequency Differential Absorption LIDAR (DIAL) System for Aerosol and Cloud Retrievals of CO2/H2O and CH4/H2O. Remote Sensing 2023, Vol 15, Page 5595 2023;15:5595. https://doi.org/10.3390/RS15235595.

[86] Yakovlev S, Sadovnikov S, Kharchenko O, Kravtsova N. Remote Sensing of Atmospheric Methane with IR OPO Lidar System. Atmosphere 2020, Vol 11, Page 70 2020;11:70. https://doi.org/10.3390/ATMOS11010070.

[87] Shayeganrad G. Single laser-based differential absorption lidar (DIAL) for remote profiling atmospheric oxygen. Opt Lasers Eng 2018;111:80–5. https://doi.org/10.1016/J.OPTLASENG.2018.07.015.

[88] Onori D, Sorianello V, Alves A, Imran M, Bogoni A, Scotti F, et al. Dual use architecture for innovative lidar and free space optical communications. Applied Optics, Vol 56, Issue 31, Pp 8811-8815 2017;56:8811–5. https://doi.org/10.1364/AO.56.008811.

[89] Wang S, Ke J, Chen S, Zheng Z, Cheng C, Tong B, et al. Performance Evaluation of Spaceborne Integrated Path Differential Absorption Lidar for Carbon Dioxide Detection at 1572 nm. Remote Sensing 2020, Vol 12, Page 2570 2020;12:2570. https://doi.org/10.3390/RS12162570.

[90] Gordon IE, Rothman LS, Hargreaves RJ, Hashemi R, Karlovets E V, Skinner FM, et al. The HITRAN2020 molecular spectroscopic database. J Quant Spectrosc Radiat Transf 2022;277:107949. https://doi.org/https://doi.org/10.1016/j.jqsrt.2021.107949.

[91] Stelman J, Liu Z. Formulation of aerial platform environmental requirements from global radiosonde data. J Unmanned Veh Syst 2017;5:62–7. https://doi.org/10.1139/JUVS-2016-0010.

[92] Fix A, Matthey R, Amediek A, Ehret G, Gruet F, Kiemle C, et al. Investigations on frequency and energy references for a space-borne integrated path differential absorption lidar. Https://DoiOrg/101117/122304145 2017;10563:74–82. https://doi.org/10.1117/12.2304145.

[93] Devara PCS, Raj PE, Pandithurai G, Dani KK, Sonbawne SM, Rao YJ. Differential absorption lidar probing of atmospheric ozone over a tropical urban station in India. Meas Sci Technol 2007;18:639. https://doi.org/10.1088/0957-0233/18/3/013.

[94] Carroll BJ, Nehrir AR, Kooi SA, Collins JE, Barton-Grimley RA, Notari A, et al. Differential absorption lidar measurements of water vapor by the High Altitude Lidar Observatory (HALO): Retrieval framework and first results. Atmos Meas Tech 2022;15:605–26. https://doi.org/10.5194/AMT-15-605-2022,.

[95] Anicho O, Belmekki BE, Arum C, Alouini M-S. High-Altitude Platform Stations (HAPS) - Bridging the Connectivity Divide and Pioneering New Frontiers: An Overview of High-Altitude Platform Stations n.d. https://www.frontiersin.org/research-topics/66725/high-altitude-platform-stations-haps---bridging-the-connectivity-divide-and-pioneering-new-frontiers-an-overview-of-high-altitude-platform-stations (accessed June 28, 2025).

[96] Services enabled by High Altitude Pseudo Satellites (HAPS) complemented by satellites n.d. https://business.esa.int/funding/invitation-to-tender/services-enabled-high-altitude-pseudo-satellites-haps-complemented-satellites (accessed June 28, 2025).

[97] Akdagl I. Exploring HAPS and Cybersecurity: The Future of High Altitude Telecommunication and Defense 2024. https://ibrahimakkdag.medium.com/exploring-haps-and-cybersecurity-the-future-of-high-altitude-telecommunication-and-defense-1ef9267ba776 (accessed June 28, 2025).

[98] Mohorcic M. CAPANINA – Communications from Aerial Platform Networks Delivering Broadband Information for All 2005.

[99] Horwath J, Knapek M, Epple B, Brechtelsbauer M, Wilkerson B. Broadband backhaul communication for stratospheric platforms: the stratospheric optical payload experiment (STROPEX). Https://DoiOrg/101117/12680824 2006;6304:436–47. https://doi.org/10.1117/12.680824.

[100] Spillard C, Grace D, Thornton J, White G, Tozer TC, Mohorcic M, et al. Broadband Communications from HeliNet High Altitude Platforms. Proceedings of DASIA 2002 – Data Systems in Aerospace, 2002.

[101] Stanley E, KayNeal D, Becker LL. Systems and methods for high-altitude radio/optical hybrid platform. US11012157B2, 2021.

[102] Sharath A, Brian B. Integrated access and backhaul from high altitude platforms. US11800374B2, 2023.

[103] Ata Y, Alouini MS. Performance of Integrated Ground-Air-Space FSO Links Over Various Turbulent Environments. IEEE Photonics J 2022;14. https://doi.org/10.1109/JPHOT.2022.3218512.

[104] Na DH, Park KH, Ko YC, Alouini MS. Beamforming and Band Allocation for Satellite and High-Altitude Platforms Cognitive Systems. IEEE Wireless Communications Letters 2022;11:2330–4. https://doi.org/10.1109/LWC.2022.3202641.

[105] Kightley P. Lidar for HAPS – Towards 3D Lidar from Near Space. QinetiQ 2021.

[106] Chouza F, Leblanc T, Wang P, Brown SS, Zuraski K, Chace W, et al. The Small Mobile Ozone Lidar (SMOL): instrument description and first results. Atmos Meas Tech 2025;18:405–19. https://doi.org/10.5194/AMT-18-405-2025,.

[107] Kallmeyer F, Strohmaier SGP, Rhee H, Hermerschmidt A, Riesbeck T, Eichler HJ, et al. Transmitter technologies for space born water vapor

[108] Galzerano G, Laporta P. Frequency-stabilized seed laser system for dial applications around 0.94 μm. Opt Lasers Eng 2006;44:677–86. https://doi.org/10.1016/J.OPTLASENG.2005.04.017.

[109] Abshire JB, Ramanathan AK, Riris H, Allan GR, Sun X, Hasselbrack WE, et al. Airborne measurements of CO2 column concentrations made with a pulsed IPDA lidar using a multiple-wavelength-locked laser and HgCdTe APD detector. Atmos Meas Tech 2018;11:2001–25. https://doi.org/10.5194/AMT-11-2001-2018,.

[110] Wührer C, Kühl C, Lucarelli S, Bode M. MERLIN: overview of the design status of the lidar instrument. International Conference on Space Optics — ICSO 2018, 2019.

[111] Qiang W, Wang C, Wang Y, Jiang Y, Li Y, Xue X, et al. All-fiber multifunction differential absorption CO 2 lidar integrating single-photon and coherent detection . Opt Express 2024;32:19665. https://doi.org/10.1364/OE.519325,.

[112] McCartney, J. E. Optics of the atmosphere: Scattering by molecules and particles. Nyjw 1976.

[113] Berman GP, Chumak AA, Gorshkov VN. Beam wandering in the atmosphere: The effect of partial coherence. Phys Rev E Stat Nonlin Soft Matter Phys 2007;76:056606. https://doi.org/10.1103/PHYSREVE.76.056606/FIGURES/5/THUMBNAIL.

[114] Wang Y, Xu H, Li D, Wang R, Jin C, Yin X, et al. Performance analysis of an adaptive optics system for free-space optics communication through atmospheric turbulence. Scientific Reports 2018 8:1 2018;8:1–11. https://doi.org/10.1038/s41598-018-19559-9.

[115] Abramov PI, Budarin AS, Kuznetsov E V., Skvortsov LA. Quantum-Cascade Lasers in Atmospheric Optical Communication Lines: Challenges and Prospects (Review). J Appl Spectrosc 2020;87:579–600. https://doi.org/10.1007/S10812-020-01041-Y/METRICS.

[116] Pang X, Ozolins O, Zhang L, Schatz R, Udalcovs A, Yu X, et al. Free-Space Communications Enabled by Quantum Cascade Lasers. Physica Status Solidi (a) 2021;218:2000407. https://doi.org/10.1002/PSSA.202000407.

[117] Pang X, Schatz R, Joharifar M, Udalcovs A, Bobrovs V, Zhang L, et al. Direct Modulation and Free-Space Transmissions of up to 6 Gbps Multilevel Signals With a 4.65-μQuantum Cascade Laser at Room Temperature. Journal of Lightwave Technology 2022;40:2370–7. https://doi.org/10.1109/JLT.2021.3137963.

[118] Vitiello MS, Scalari G, Williams B, Natale P De. Quantum cascade lasers: 20 years of challenges. Opt Express 2015;23:5167–82. https://doi.org/10.1364/OE.23.005167.

[119] Mhanna M, Sy M, Elkhazraji A, Farooq A. A laser-based sensor for selective detection of benzene, acetylene, and carbon dioxide in the fingerprint region. Appl Phys B 2023;129:1–9. https://doi.org/10.1007/S00340-023-08083-Y/FIGURES/7.

[120] Elkhazraji A, Sy M, Mhanna M, Aldhawyan J, Khaled Shakfa M, Farooq A. Selective BTEX detection using laser absorption spectroscopy in the CH bending mode region. Exp Therm Fluid Sci 2024;151:111090. https://doi.org/10.1016/J.EXPTHERMFLUSCI.2023.111090.

[121] Elkhazraji A, Shakfa MK, Shakfa MK, Abualsaud N, Mhanna M, Sy M, et al. Laser-based sensing in the long-wavelength mid-infrared: chemical kinetics and environmental monitoring applications. Applied Optics, Vol 62, Issue 6, Pp A46–A58 2023;62:A46–58. https://doi.org/10.1364/AO.481281.

[122] Elkhazraji A, Khaled Shakfa M, Lamperti M, Hakimov K, Djebbi K, Gotti R, et al. High-resolution molecular fingerprinting in the 11.6–15 μm range by a quasi-cw difference-frequency-generation laser source. Opt Express 2023. https://doi.org/https://doi.org/10.48550/arXiv.2212.09089.

[123] Elkhazraji A, Sy M, Shakfa MK, Farooq A. A mid-infrared laser diagnostic for simultaneous detection of furan and nitric oxide. Proceedings of the Combustion Institute 2024;40:105366. https://doi.org/10.1016/J.PROCI.2024.105366.

[124] Shakfa MK, Lamperti M, Gotti R, Gatti D, Elkhazraji A, Hakimov K, et al. A widely tunable difference-frequency-generation laser source for high-resolution spectroscopy in the 667–865 cm−1 range. Proc.SPIE, vol. 11670, 2021. https://doi.org/10.1117/12.2577534.

[125] Hu P, Zhang J, Yoon M, Qiao XF, Zhang X, Feng W, et al. Highly sensitive phototransistors based on two-dimensional GaTe nanosheets

with direct bandgap. Nano Res 2014;7:694–703. https://doi.org/10.1007/S12274-014-0430-2/METRICS.

[126] Alouini M-S, Ooi BS, Trichili A, Sait M, Ng TK, Alshaibaini S, et al. Dual-wavelength luminescent fibers receiver for wide field-of-view, Gb/s underwater optical wireless communication. Optics Express, Vol 29, Issue 23, Pp 38014-38026 2021;29:38014–26. https://doi.org/10.1364/OE.443255.

[127] Kong M, Sun X, Ng TK, Ooi BS, Alfaraj N, Sait M, et al. The effect of turbulence on NLOS underwater wireless optical communication channels [Invited]. Chinese Optics Letters, Vol 17, Issue 10, Pp 100013-2019;17:100013-.

[128] Arumugam M. Optical fiber communication—An overview. Pramana 2001 57:5 2001;57:849–69. https://doi.org/10.1007/S12043-001-0003-2.

[129] Nguyen Van H, Loghmari Z, Philip H, Bahriz M, Baranov AN, Teissier R. Long Wavelength (λ >17 μm) Distributed Feedback Quantum Cascade Lasers Operating in a Continuous Wave at Room Temperature. Photonics 2019;6. https://doi.org/10.3390/photonics6010031.

[130] Makarov DS, Tretyakov MY, Rosenkranz PW. Revision of the 60-GHz atmospheric oxygen absorption band models for practical use. J Quant Spectrosc Radiat Transf 2020;243:106798. https://doi.org/10.1016/J.JQSRT.2019.106798.

[131] Mhanna M, Sy M, Elkhazraji A, Farooq A. Deep neural networks for simultaneous BTEX sensing at high temperatures. Optics Express, Vol 30, Issue 21, Pp 38550-38563 2022;30:38550–63. https://doi.org/10.1364/OE.473067.

[132] Mhanna M, Sy M, Elkhazraji A, Farooq A. Multi-Speciation in Shock Tube Kinetics using Deep Neural Networks and Cavity-Enhanced Absorption Spectroscopy. Proceedings of the Combustion Institute 2024;40. https://doi.org/https://doi.org/10.1016/J.PROCI.2024.105395.

[133] Mhanna M, Sy M, Elkhazraji A, Farooq A. Multi-Species Sensing Using CEAS and DNN in Shock Tube Kinetics. In Computational Optical Sensing and Imaging (pp. JTu2A-4), Optica Publishing Group; 2022.

[134] Mhanna M, Sy M, Elkhazraji A, Farooq A. Laser-based sensor for multi-species detection using ceas and dnn. Conference on Lasers and Electro-Optics (CLEO), IEEE; 2022.

[135] Barton-Grimley RA, Nehrir AR, Kooi SA, Collins JE, Harper DB, Notari A, et al. Evaluation of the High Altitude Lidar Observatory (HALO) methane retrievals during the summer 2019 ACT-America campaign. Atmos Meas Tech 2022;15:4623–50. https://doi.org/10.5194/AMT-15-4623-2022,.